# On the Relationship between BL-Lacertae Objects and FRI Radio Galaxies

Geoffrey V. Bicknell
Mt. Stromlo and Siding Spring Observatories
Institute of Advanced Studies
Australian National University

**Abstract**

The idea that Fanaroff-Riley Class I Radio Galaxies contain core jets with Lorentz factors of the order of a few and are the parent population for BL Lac Objects is examined, with particular reference to the data on two FRI Radio Galaxies, NGC 315 and NGC 6251. Conservation laws for an entraining relativistic jet are derived and are used to determine a relationship between Mach number and velocity for initially relativistic jets. One immediate consequence of this relationship is that, if an initially relativistic jet is decelerated to a transonic Mach number between, say 0.5 and 1.5, then its velocity at this point is between 0.3 and 0.7 times the speed of light. Analysis of the milliarcsecond and arcsecond data on NGC 315 and NGC 6251 shows that the Lorentz factors of the core jets in these galaxies can be as high as 2-4 provided that the jet pressures are not much more than an order of magnitude higher than the minimum pressures. Such Lorentz factors are more plausible if the jets are inclined at $\sim$ 30 degrees to the line of sight, although larger angles would also be possible. For core Lorentz factors in the range 2-4, the jets are mildly relativistic on the large scale and their velocities are consistent with the jet brightness asymmetries being caused by Doppler beaming. The Lorentz factors required for significant beaming of core jets may be reconciled with subluminal motions of knots in the core if these are reverse shocks advected by the jet. Moderate jet Lorentz factors are possible if the shocks are normal and higher Lorentz factors are possible if the shocks are oblique. This paper provides support for the idea that the transition from two-sided to one-sided jets across the FRI/II break is due to the transition from subrelativistic to relativistic flow.

# Contents

# 1 Introduction

There are strong (and well known) arguments as to why the jets in the cores of various forms of active galaxies are moving relativistically. The reconciliation of the time scales of variation with brightness temperatures (Rees, 1966), superluminal motion (see, for example, the review by Cawthorne (1991)) and the explanation for the low level of core X-ray emission (Burbidge, Jones and O'Dell, 1974) are the main reasons why the idea that there exists relativistic bulk motion, in both quasars and BL-Lac objects has been favoured for some time. More recently, relativistic core jets have come to play a strong role in *Unified Schemes* relating quasars and Fanaroff-Riley (1974) Class II (FRII) radio galaxies (e.g. Barthel 1989) as well as BL-Lac objects and Fanaroff-Riley Class I (FRI) radio galaxies (Blandford and Rees 1978, Padovani and Urry 1990). The essential ingredient of these schemes is that Doppler-beamed emission from the core is enhanced when a jet is viewed at a small to moderate angle to the line of sight and the ratio of core to extended flux density is boosted considerably in such cases. In the case of BL-Lacs the optical luminosity function of the parent galaxy and the extended radio emission implicate FRI radio galaxies as the parent population (Ulrich, 1989, Ulvestad and Antonucci, 1986). However, it is interesting that most of the BL Lacs in the Ulrich sample are in the range of optical and radio luminosities approximately a decade below the FRI/II break. Thus they appear to represent the *more luminous* FRI's. While there is a strong body of opinion favoring this idea, there are some differences in the range of Lorentz factors thought to be relevant. Padovani and Urry (1990) in fitting the luminosity functions of BL Lacs in various bands require Lorentz factors $\approx 3$ for X-ray selected BL-Lacs, whereas optical and radio-selected BL-Lacs, in their scheme, require Lorentz factors $\sim 7-20$. Hughes, Aller and Aller (1989b) used a Lorentz factor of 4 in their model of the flux variations in BL-Lac itself. Mutel (1992) has pointed out that VLBI observations of BL-Lac objects usually indicate $2 < \gamma_{\rm radio} < 4$. He also noted that there is no theory which predicts an acceleration between the X-ray and optical-radio emitting regions. The reconciliation between the different ranges of Lorentz factors may involve more detailed consideration of pattern versus beam speeds (Lind and Blandford, 1985, Cohen and Vermeulen, 1992). However, despite these differences, it seems certain that unification of BL-Lacs with FRI radio sources involves pc-scale Lorentz factors of at least a few.

Unification of BL-Lac objects and FRI galaxies poses some interesting questions: Estimates of FRI jet velocities on the kiloparsec scale (e.g. Bicknell 1986a,b, Bicknell *et al.*1990) are typically of the order of $1,000 - 10,000\,{\rm km\,s^{-1}}$. Although in the latter paper, model dependent estimates of density and Mach number were used to estimate jet velocities, one can also use a (more or less) model-independent estimate based on the energy flux (Bicknell, 1986a, see also § 5). Provided that one has a reliable estimate of the age of the lobe it is possible to estimate the velocity at different points along the jet (or at least a maximum value). Frequently this again gives velocities for class I jets of the order of $1,000 - 10,000\,{\rm km\,s^{-1}}$. Thus, consistency with FRI/BL Lac unification requires the initially relativistic jets to have been decelerated somewhere between the pc and kpc scale. This raises the questions as to where the deceleration takes place and whether the required deceleration is dynamically consistent.

Another issue which inevitably arises when discussing relativistic jet flow is the explanation for jet sidedness. A common notion is that two-sided jets are subrelativistic and one-sided jets are relativistic and that the transition from two-sided to one-sided jets near the FRI/II break (Bridle, 1984) represents a transition from subrelativistic to relativistic flow. In addition, the bases of two-sided jets tend to be one-sided (Bridle and Perley, 1984) suggesting that the bases of even FRI jets are at least mildly relativistic. This view has been developed by Laing (1993) in a model for the polarization structure of the bases of low-powered jets.

The purpose of this paper is to examine some of the physics relating to deceleration of an initially relativistic parsec-scale jet to transonic, subrelativistic or mildly relativistic kpc-scale flow. The main approach is to use conservation laws based on energy and momentum to ascertain downstream parameters given the initial Lorentz factor and surface brightnesses and jet diameters on both the parsec and kiloparsec scale. The conservation laws and some immediate deductions from them are given in § 3. The related physics of jet beaming, sidedness and subluminal motion are discussed in § 4. The approach used in § 3 is similar to that employed in Bicknell (1986b). That is, the jet conservation laws are used without a detailed specification of the turbulent stresses which describe the entrainment process in detail. The rationale for this approach is that there is no secure description for "turbulent viscosity" in the case of supersonic non-relativistic jets and any prescription for relativistic jets would necessarily be highly controversial. On the other hand, the combination of energy and momentum conservation plus jet data allow one to determine the kpc-scale velocity and Mach number and the density of entrained matter, given initial conditions and some simplifying assumptions.

The main scenario envisaged is the following: An initially free relativistic jet comes into static pressure equilibrium with the interstellar medium (ISM) of the parent galaxy at a distance of the order of a 100 pc. The jet starts to interact with the ISM at this point, entraining thermal matter causing it to decelerate to a Mach number near unity, leading to the onset of fully-developed turbulence and consequent rapid spreading. This scenario is qualitatively supported by the observations of at least two relatively nearby FRI galaxies, IC4296 (Killeen, Bicknell and Ekers, 1986) and NGC 315 (Bridle, 1982; Venturi *et al.*, 1993) in which thin knotty jets start to rapidly expand a few kpc from the nucleus.

A large part of the analysis presented in this paper is based upon two Radio Galaxies, NGC 315 and NGC 6251. For both of these galaxies good radio data are available ranging from the milliarcsecond to the arcminute scales. The conditions under which entrainment is sufficient to decelerate relativistic jets in NGC 315 and NGC 6251 to transonic, mildly relativistic velocities are derived in §§ 6 and 7 using energy and momentum conservation in combination with the data on both the pc-scale and kpc-scale jets. Two galaxies, of course, do not constitute a statistically significant sample. However, the observations of these two radio sources has played a prominent rôle in the discussion of relativistic effects in the past, and therefore it is appropriate that they be the first to be examined in a detailed fashion with respect to the above-mentioned ideas. As comparable VLBI data become available for other well-studied radio galaxies, it will be possible to extend this analysis to a larger sample. The computer programs used in this paper will be made readily available for this purpose.

# 2 The Equations for Relativistic Flow in a Gravitational Field

## 2.1 Fundamental equations

When considering relativistic flow it is not usually necessary to consider the effect of the gravitational field since the plasma is light and fast. In considering the conservation of momentum in an entraining jet, however, we need to take into account the effect of the external pressure gradient and the transition of the jet density to that of the background. As we shall see considerations of these effects involve the gravitational field. Thus, although the rôle of the gravitational field in the conservation laws is minor in comparison to the complication of the relativistic hydrodynamic equations involved in introducing it, its incorporation is logically necessary. Hence I consider a *fast motion - weak field* approximation for relativistic flow in a gravitational field. The relevant equations can be derived from the general relativistic equations for fluid flow by an expansion to first order in $\phi/c^2$ where $\phi$ is the Newtonian gravitational potential. In the following the dynamical effect of the magnetic field is largely ignored, except when considering the energy budget in the lobe. This is justifiable if the fluid particle pressure dominates the magnetic pressure and is an acceptable approach to take for a "first pass" at this problem. If the magnetic field is tangled, it may be incorporated in an approximate fashion as an additional pressure (albeit with a different equation of state). Thus, even of the magnetic pressure is important, the calculations reported here probably still provide a reasonable guide to the actual situation.

To first order in $\phi/c^2$, the metric is:

$$ds^2 = g_{ij} dx^i dx^j = \left(1 + 2\frac{\phi}{c^2}\right)(dx^0)^2 - \left(1 - 2\frac{\phi}{c^2}\right)(d\mathbf{x})^2 \tag{1}$$

($x^0 = ct, \mathbf{x}$) is the background Minkowskian coordinate system (Landau & Lifshitz, 1975, p338). In the following, the Landau & Lifshitz convention for numbering indices is followed.

The fluid equations are most easily evaluated from the conservation of the stress energy tensor $T^{ij} = w u^i u^j - p g^{ij}$ ($w = e + p$ is the relativistic enthalpy, $e$ is the internal energy, $p$ is the pressure, and $u^i = dx^i/ds$ are the components of the four-velocity). The conservation equations are most conveniently expressed in the following form:

$$T^j_{i;j} = \frac{1}{\sqrt{-g}} \frac{\partial}{\partial x^j} \left(\sqrt{-g}\, T^j_i\right) - \frac{1}{2} \frac{\partial g_{kl}}{\partial x^i} T^{kl} = 0 \tag{2}$$

resulting in the following equations:

Energy

$$\frac{\partial}{\partial t}\left[u^{0^2} w - p\left(1 - 2\frac{\phi}{c^2}\right)\right] + \frac{\partial}{\partial x^\alpha}\left[u^{0^2} w\, v^\alpha\right] = 0 \tag{3}$$

Momentum

$$\frac{\partial}{\partial t}\left[\left(1 - 4\frac{\phi}{c^2}\right) u^{0^2} \frac{w}{c^2} v^\alpha\right] + \frac{\partial}{\partial x^\beta}\left[\left(1 - 4\frac{\phi}{c^2}\right) u^{0^2} \frac{w}{c^2} v^\alpha v^\beta\right]$$
$$+ \left(1 - 2\frac{\phi}{c^2}\right) \frac{\partial p}{\partial x^\alpha} + u^{0^2} \frac{w}{c^2} \left(1 + \frac{v^2}{c^2}\right) \frac{\partial \phi}{\partial x^\alpha} = 0 \tag{4}$$

where $v^\alpha/c = dx^\alpha/dx^0$ and $u^0 = dx^0/ds$ satisfies, to first order in $\phi/c^2$,

$$u^{0\,2} = \gamma^2 \left[1 - \frac{2\phi}{c^2}(2\gamma^2 - 1)\right] \tag{5}$$

where $\gamma = (1 - v^2/c^2)^{-1/2}$ is the Lorentz factor of the fluid in the background Minkowski space time. (The term in $v^2/c^2$ in the last term in this equation is at first surprising. However, its origin is in the 3-space Christoffel symbols and when this equation is expressed in the curved 3-space, orthogonal to $x^0$, the $v^2/c^2$ is absorbed into the curved 3-space divergence operator. This reflects the fact that it is the curved space time which is strictly observable not the background Minkowskian system.)

Particle Number

Conservation of particle number is expressed by:

$$\frac{1}{\sqrt{-g}}\frac{\partial}{\partial x^i}\left(\sqrt{-g}\, n\, u^i\right) = 0 \tag{6}$$

where $n$ is the proper particle number density of any conserved species. To first order in $\phi/c^2$:

$$\frac{\partial}{\partial t}\left[\left(1 - 2\frac{\phi}{c^2}\right) u^0 n\right] + \frac{\partial}{\partial x^\alpha}\left[\left(1 - 2\frac{\phi}{c^2}\right) u^0 n\, v^\alpha\right] = 0 \tag{7}$$

This equation formally shows that the correspondence between the proper particle density and the particle density in the "lab" frame is given by $n_{\text{lab}} = (1 - 2\phi/c^2)\, u^0\, n$, generalizing the special relativistic result $n_{\text{lab}} = \gamma\, n$. This relation is important in deriving the correct non-relativistic limit of the above equations. This is straightforward and is therefore not given here.

# 3  Conservation Laws for a Relativistic Entraining Jet

As argued by Begelman, Blandford and Rees (1984) and specifically for NGC 6251 by D.L. Jones *et al.* (1986), it is unlikely that core jets are confined by the ISM. However, the pressure in the jet decreases as (Jet Radius)$^{-8/3}$. For the FRI galaxies considered here typical values of the core jet pressure (dependent upon beaming) are $\sim 10^{-5}$ dy cm$^{-2}$ while typical central values of the ISM pressure (corresponding to a number density $\sim 1$ cm$^{-3}$ and $T_{ISM} \sim 10^7$ $K$) are $\sim 10^{-9}$ dy cm$^{-2}$. Thus when the jet has expanded by approximately a factor of 30, that is to a diameter of about 30 pc, it will start to some into pressure equilibrium with the ISM. Assuming a nominal expansion rate of $\sim 0.1$ implies that this will occur at the order of 300 pc from the core. This is the order of magnitude of the gap region observed in a number of relatively nearby FRI galaxies. Thus the main scenario considered in this paper is that the jets are in free expansion until they start to interact with the ISM at the order of 100 pc from the core and this interaction initiates entrainment which decelerates the jet. This scenario guides to some extent the final form and the application of the conservation laws derived in this section. However, the relationships derived here are also relevant to other scenarios. For example, one could envisage an increasing pressure within the optical core radius of the galaxy due to the central object so that the jet entrains this atmosphere all the way out from the core.

Let us now consider the appropriate integral forms of the fluid dynamic equations derived in the preceding section. For the sake of simplicity let us assume that the pressure external to the jet, $p_{\text{ext}}$ is spherically symmetric. The following conservation laws are obtained by integrating throughout the control volume $V$ bounded by the control surface $S$ which, because of the assumed spherical symmetry of the background medium, is "capped" by sections of constant spherical radius and whose side, $S_E$ (the entrainment surface) is situated a large enough distance from the jet that external conditions apply (see figure 1). It is assumed that the flow is time independent (at least in an average sense). Initial (pc-scale) flow variables are denoted by a '1' subscript; final (kpc-scale) flow variables are denoted by a '2' subscript and it is assumed that no matter what the final velocity of the final state, the internal energy and pressure are nevertheless dominated by relativistic particles. For the sources we consider, this condition will be shown to be valid. The jet propagates in the direction of the $z$-axis and the distinction between the $z$ and $r$-components of velocity and the magnitude of the velocity across the jet is ignored. The cross-sectional area of the jet is denoted by $A$. (Note that $r$ refers to the spherical radial coordinate.)

In the following, most of the terms in $\phi/c^2$ are ignored, save for the gravitational term in the momentum equation. That this can be done appears obvious by inspection of the equations in the previous section, since $\phi/c^2 << 1$. The neglect of the potential terms in the energy equation is not quite so obvious since, to recover the non-relativistic equations, one expands to terms in order $1/c^2$ to arrive at the relevant form of the energy equation. (The first order terms correspond to conservation of mass.) However, one can neglect the potential when $v_{\text{jet}}^2 >> \Delta\phi$ where $\Delta\phi$ is the difference in the potential along the jet. The reason for keeping the gravitational term in the momentum equation will become apparent in what follows.

## 3.1 Conservation of particle number and rest-mass

Integration of the particle conservation equation (equation (7)) throughout $V$ gives the following conservation law for species which are neither created nor annihilated in $V$.

$$\gamma_2 n_2 \, v_2 \, A_2 = \gamma_1 \, n_1 \, v_1 \, A_1 - \int_{S_E} n_{\text{ISM}} \, \mathbf{v}_{\text{ent}} \cdot \mathbf{n} \, dS \tag{8}$$

Thus, considering thermal material,

$$\gamma_2 \, \rho_2 \, v_2 \, A_2 = \gamma_1 \, \rho_1 \, v_1 \, A_1 - \int_{S_E} \rho \, \mathbf{v}_{\text{ent}} \cdot \mathbf{n} \, dS \tag{9}$$

where $\rho = m \, n$ is the rest-mass density and $m \approx 0.59 \, m_p$ is the mean mass per particle of the thermal plasma. Note that I am assuming that the flow on $S_E$ is non-relativistic. The existence of polarization in many BL Lac VLBI jets (Roberts et al., 1990) indicates that they initially contain very little thermal plasma so that the bulk of the thermal plasma results from entrainment. Therefore,

$$\gamma_2 \, \rho_2 \, v_2 \, A_2 \approx - \int_{S_E} \rho \, \mathbf{v}_{\text{ent}} \cdot \mathbf{n} \, dS \tag{10}$$

If the final flow is non-relativistic, then this equation is simply:

$$\rho_2 \, v_2 \, A_2 \approx - \int_{S_E} \rho \, \mathbf{v}_{\text{ent}} \cdot \mathbf{n} \, dS \tag{11}$$

## 3.2 Conservation of momentum

### 3.2.1 General relations

The flux of the $\alpha$ component of momentum is $\gamma^2 \, (w/c^2) \, v^\alpha \, v^\beta + p \delta^{\alpha\beta}$. However, the momentum flux associated with the jet, $(\gamma^2 \, (w/c^2) \, v^2 + p) \, A$ is *not* conserved because of the effect of the background pressure. Mathematically, this manifests itself through the flux, in the jet direction, of the external pressure, integrated over the sides of the control volume. In order to correctly determine the correct expression of the conservation of momentum one writes equation (4) in the form (for time independent flow):

$$\frac{\partial}{\partial x^\beta} \left( \frac{w}{c^2} v^\alpha \, v^\beta \right) = - \frac{\partial p}{\partial x^\alpha} - \rho^* \frac{\partial \phi}{\partial x^\alpha} \tag{12}$$

where $\rho^* = \gamma^2 \, w/c^2 \, (1 + v^2/c^2)$. Since the background medium is spherically symmetric the external pressure $p_{\text{ext}}(r)$ external density $\rho_{\text{ext}}(r)$ and the potential $\phi(r)$ are related by

$$\frac{dp_{\text{ext}}}{dr} = -\rho_{\text{ext}} \frac{d\phi}{dr} \tag{13}$$

Hence, equation (12) may be written

$$\frac{\partial}{\partial x^\beta} \left[ \gamma^2 \frac{w}{c^2} v^\alpha \, v^\beta + \Delta p \, \delta^\alpha_\beta \right] = - \left( 1 - \frac{\rho^*}{\rho_{\text{ext}}} \right) \frac{dp_{\text{ext}}}{dr} \frac{x^\alpha}{r} \tag{14}$$

where $\Delta p = p - p_{\text{ext}}(r)$ is the difference between the internal and external pressures at the same radius. One can now integrate this equation throughout the control volume indicated in figure 1 with the result:

$$\int_{S_2} \left[ \gamma^2 \frac{w}{c^2} v^\alpha v^r + \Delta p \right] dS - \int_{S_1} \left[ \gamma^2 \frac{w}{c^2} v^\alpha v^r + \Delta p \right] dS + \int_{S_E} \rho v^\alpha \mathbf{v}_{\text{ent}} \cdot \mathbf{n} \, dS$$
$$= - \int \frac{dp_{\text{ext}}}{dr} \left( 1 - \frac{\rho^*}{\rho_{\text{ext}}} \right) \frac{x^\alpha}{r} dV \qquad (15)$$

Note that $\Delta p = 0$ on $S_E$ simplifies the integral over that surface. The integral over $S_E$ represents the entrainment of momentum which one would expect, on physical grounds, to be zero. Mathematically the argument is as follows: Since, in general, a finite amount of matter may be entrained, the integral of $\rho \mathbf{v}_{\text{ent}} \cdot \mathbf{n}$ over $S_E$ is non-zero. However, when this quantity is multiplied by the vanishingly small component of velocity, the contribution of the integral over $S_E$ on the left hand side, vanishes.

Taking the z-component of equation (15), with $v^z \approx v^r \approx v_{\text{jet}}$

$$(\gamma_2^2 \frac{w}{c^2} v_2^2 + \Delta p_2) A_2 \approx \left( \gamma_1^2 \frac{w}{c^2} v_1^2 + \Delta p_1 \right) A_1 - \int_{r_1}^{r_2} dr \left[ \frac{dp_{\text{ext}}}{dr} \int_A \left( 1 - \frac{\rho^*}{\rho_{\text{ext}}} \right) \frac{z}{r} dS \right] \qquad (16)$$

One can now appreciate the reason for the introduction of the gravitational field. Referring to equation (16), note that since the density $\rho^* \to \rho_{\text{ext}}$ outside the jet, the contribution of the integral on the right hand side is finite. If the gravitational term in the momentum equation had not been included then this integral would have had an infinite contribution, physically corresponding to driving of the external medium outwards by the external pressure gradient. Another approach to this problem could involve neglecting the gravitational field in the relativistic portion of the flow and including its effect in the non-relativistic outer portion. However, such an approach would be clumsy and not as elegant as including the gravitational field in a consistent fashion from the outset.

Note also that the conservation of momentum is expressed in terms of an integral involving $\Delta p$ and that the external contribution to the momentum flux involves the density differential. The term on the right of equation (16) represents the net effect of buoyancy on the jet. For a confined jet this is particularly important for low Mach numbers. In the scenario under consideration here, the term is unimportant, since, for most of the jet region from the core to the kpc scale, $p_{\text{ext}} \ll p_{\text{jet}}$. Moreover, if there is no increase in the ISM pressure due to the central object, then $dp_{\text{ext}}/dr \approx 0$ within a core radius of the optical galaxy. The length scale, $L$, of the variation in the momentum of a confined jet, compared to the local pressure scale height, $h$, can be determined from equation 16 to be:

$$\frac{L}{h} \approx 2 \, \mathcal{M}^2 \qquad (17)$$

Thus for high Mach numbers the buoyant driving force is also unimportant.

For a jet which is initially free and which then becomes confined by the ISM, $\Delta p_1 \approx p_1$ and $\Delta p_2 \approx 0$. Hence,

$$\gamma_2^2 \frac{w_2}{c^2} v_2^2 A_2 \approx \left( \gamma_1^2 \frac{w_1}{c^2} v_1^2 + p_1 \right) A_1 \qquad (18)$$

Splitting the relativistic enthalpy into components of rest-mass energy density ($\rho \, c^2$), internal energy density ($\epsilon$) and pressure ($p$) ($w = \rho \, c^2 + \epsilon + p$) and introducing the ratio $\mathcal{R} = (\rho \, c^2)/(\epsilon + p)$ of rest-mass energy to enthalpy, equation (16) may be written in the form:

$$\gamma_2^2 \beta_2^2 (1 + \mathcal{R}) \approx \frac{1 + 4\gamma_1^2 \beta_1^2}{4} \left( \frac{p_1 A_1}{p_2 A_2} \right) \tag{19}$$

where $\beta = v/c$ and $p = \epsilon/3$.

The transition from a relativistically dominated jet to one in which the inertia is dominated by thermal matter, is signaled by the aproach of $\mathcal{R}$ to unity. This parameter also has the following significance. The ratio of jet to ISM density $\eta = \rho/\rho_{ISM}$ can be expressed in the form:

$$\begin{aligned} \eta &= \mathcal{R} \left( \frac{p_{\text{jet}}}{p_{\text{ISM}}} \right) \left( \frac{4kT}{\mu m_p c^2} \right) \\ &= 6.1 \times 10^{-6} \, \mathcal{R} \left( \frac{p_{\text{jet}}}{p_{\text{ISM}}} \right) \left( \frac{T_{\text{ISM}}}{10^7 \, K} \right) \end{aligned} \tag{20}$$

This expression shows that the jet inertia needs to be very strongly dominated by thermal particles before it is of the same density as the external medium, that is we require $\mathcal{R} \sim 10^5$, before this can happen. Thus the assumption of a relativistic equation of state relating pressure and the internal energy density is warranted in all regions of the flow.

Another way of looking at this relationship, is that $\mathcal{R} = 1$ identifies a critical jet density ratio

$$\eta_{\text{crit}} = 6.1 \times 10^{-6} \left( \frac{p_{\text{jet}}}{p_{\text{ISM}}} \right) \left( \frac{T_{\text{ISM}}}{10^7 \, K} \right) \tag{21}$$

above which the jet inertia becomes dominated by thermal particles. (In terms of this definition $\mathcal{R} = \eta/\eta_{\text{crit}}$.)

The parameter $\mathcal{R}$ also appears in the expressions for the jet sound speed $c_s$ and Mach number $\mathcal{M}$:

$$\frac{c_s^2}{c^2} = \frac{1}{3(1 + \mathcal{R})} \tag{22}$$

$$\mathcal{M}^2 = (2 + 3\mathcal{R})\gamma^2 \beta^2 \tag{23}$$

where Königl's (1980) definition of relativistic Mach number ($\mathcal{M} = \gamma_v v/\gamma_{c_s} c_s$) has been used. (Königl showed that this definition is the appropriate relativistic generalization of the Newtonian expression.) It is easily shown that equation (23) reduces to the usual expression in the non-relativistic limit.

### 3.2.2 Momentum conservation in the nonrelativistic limit

When the final flow is non-relativistic equation (19) simplifies to:

$$\rho \, v_2^2 \, A_2 \approx \left( \gamma_1^2 \frac{w_1}{c^2} v_1^2 + p_1 \right) \tag{24}$$

and the final Mach number can be expressed in the form:

$$\mathcal{M}_2^2 = \frac{3}{4} \left( 4 \gamma_1^2 \beta_1^2 + 1 \right) \left( \frac{p_1 A_1}{p_2 A_2} \right) \tag{25}$$

## 3.3 Conservation of Energy

### 3.3.1 General relations

The integration of the energy equation (3) over the control volume yields:

$$\gamma_2^2 \, w_2 \, v_2 \, A_2 = \gamma_1^2 \, w_1 \, v_1 \, A_1 - \int_{S_E} \left( \rho \, c^2 + \epsilon + p \right) \mathbf{v}_{\text{ent}} \cdot \mathbf{n} \, dS \qquad (26)$$

The relativistic energy flux incorporates the rest-mass energy flux. Moreover, the entrainment of rest-mass energy is incorporated into the integral on the right hand side. These terms may be eliminated using equation (9) for the conservation of rest-mass to give:

$$\left[ (\gamma_2^2 - \gamma_2) \rho_2 \, c^2 + 4 \, \gamma_2^2 \, p_2 \right] v_2 \, A_2 = 4 \, \gamma_1^2 \, p_1 \, v_1 \, A_1 - \int_{S_E} \rho \, h \, \mathbf{v}_{\text{ent}} \cdot \mathbf{n} \, dS \qquad (27)$$

where the rest-mass flux in the parsec scale jet has been ignored and the non-relativistic specific enthalpy of the ISM, $h = \rho^{-1}(\epsilon + p)$. The left hand side of this equation constitutes the jet energy flux which consists of terms describing the flux of total energy subtracted from which is the term, $\gamma_2 \, \rho_2 \, c^2 \, v_2 \, A_2$, corresponding to the flux of rest-mass energy.

The entrainment integral now contains terms corresponding to the addition of "low-grade" enthalpy to the jet and, in the region of the jet in which we are interested, this can be safely ignored. More specifically, the entrainment term in equation (27) can be written as:

$$F_{\text{E,ent}} = -h_{\text{ISM}} \int \rho \, \mathbf{v}_{\text{ent}} \cdot \mathbf{n} \, dS = h_{\text{ISM}} \, (F_{\text{M},2} - F_{\text{M},1}) \qquad (28)$$

where $F_{\text{M}}$ is the mass flux and $h_{\text{ISM}}$ is the specific enthalpy ($= 2.5(\mu \, m_p)^{-1} \, kT$) of the interstellar medium. For an isothermal medium the specific enthalpy is constant; when it is not constant $h_{\text{ISM}}$ should be understood as an "entrainment averaged" value. For the region of the jet under consideration $F_{M,1} << F_{M,2}$. The ratio of the entrainment energy flux to the energy flux at position 2, is:

$$\frac{F_{\text{E,ent}}}{F_{E,2}} = \frac{5}{8} \frac{\eta}{\gamma_2 + (\gamma_2 - 1) \mathcal{R}_2} \qquad (29)$$

where $\eta$ is the jet density ratio, $\mathcal{R}$ is the previously introduced ratio of mass-energy density to enthalpy and transverse pressure equilibrium has been assumed. Thus, in the region of the jet in which we are interested in this section, which is still light ($\eta << 1$), the entrainment energy flux is insignificant compared to the total energy flux. In the non-relativistic limit this ratio becomes $\frac{5}{8} \eta (1 + \mathcal{M}^2/6)^{-1}$. Thus in all regions of the jet, the entrainment energy flux is unimportant as long as the jet is light.

Another useful relationship is that between the mass and energy fluxes:

$$\begin{aligned} F_{\text{M}} &= \frac{1}{4} \frac{\eta}{1 + (\gamma_2 - 1)\mathcal{R}} \frac{\mu \, m_p}{kT_{\text{ISM}}} F_{\text{E}} \\ &= 3 \frac{\eta}{1 + (\gamma_2 - 1)\mathcal{R}} \left( \frac{F_{\text{E}}}{10^{42} \text{ergs s}^{-1}} \right) \left( \frac{T_{\text{ISM}}}{10^7 \, K} \right)^{-1} M_\odot \, \text{y}^{-1} \end{aligned} \qquad (30)$$

Thus, neglecting the entrainment of energy, conservation of energy can be expressed in the form:

$$\left[ (\gamma_2 - 1) \, \mathcal{R} + \gamma_2 \right] \gamma_2 \, \beta_2 = \gamma_1^2 \, \beta_1 \left( \frac{p_1 \, A_1}{p_2 \, A_2} \right) \qquad (31)$$

This equation may be used in conjunction with the momentum equation (19) to solve for both $\mathcal{R}$ and $\beta_2$. The equations constitute a non-linear set so that a non-linear root finding algorithm is required. The subroutine *mnewt* from *Numerical Recipes* (Press et al., 1986) is quite satisfactory for this purpose.

### 3.3.2 Energy conservation in the non-relativistic limit

The above equation (31)) reduces to the following when the final flow is non-relativistic:

$$\left(\frac{1}{2} \rho_2 v_2^2 + \epsilon_2 + p_2\right) v_2 A_2 = \gamma_1^2 w_1 v_1 A_1 \tag{32}$$

giving

$$\beta_2 = \frac{\gamma_1^2 \beta_1}{1 + \frac{\mathcal{M}^2}{6}} \left(\frac{p_1 A_1}{p_2 A_2}\right) \tag{33}$$

This equation can be used with equation (25) for the non-relativistic Mach number when the final flow is non-relativistic, to solve for both $\beta_2$ and the Mach number. Unlike the relativistic case, the solution is quite straightforward. For a nonrelativistic final state the two independent methods of solution provide a useful cross-check.

## 3.4 Immediate consequences of energy and momentum conservation

The same factor of $(p_1 A_1)/(p_2 A_2)$ occurs in the expressions derived from energy and momentum conservation. Using equations (19) and (31) one finds that $\mathcal{R}$ is related to $\beta_2$ in the following way:

$$\mathcal{R} = \frac{\gamma_2 (1 - X \beta_2)}{1 - \gamma_2 (1 - X \beta_2)} \tag{34}$$

$$\text{where} \quad X(\beta_1) = \frac{4 \gamma_1^2 \beta_1}{1 + 4 \gamma_1^2 \beta_1^2} \tag{35}$$

For a jet which is initially ultrarelativistic $X \approx 1$ and in fact $X(\beta)$ does not vary by more than $\pm 15\%$ over the range $0.25 < \beta < 1$. The value of $\mathcal{R}$ as a function of $\beta_2$ is plotted in the left hand panel of figure 2 for a number of values of $\gamma_1$. The Mach numbers corresponding to the above expression for $\mathcal{R}$ are also plotted in the right hand panel of figure 2. The two most interesting points about these plots are that the Mach number becomes subsonic when $\beta_2 \approx 0.3$ and, moreover, this is the point where the jet inertia is becoming thermally dominated with values of $\mathcal{R} \approx 2 - 4$.

The existence of a critical velocity $\approx 0.3\,c$ has a simple physical interpretation: The ratio of energy to momentum fluxes in an ultrarelativistic jet is $c$; for subrelativistic flow, the ratio of energy to momentum fluxes is $3v(1 + \mathcal{M}^2/6)\mathcal{M}^{-2}$. Equating the two gives, for $\mathcal{M} = \infty$, $v/c = 2/7 \approx 0.29$. This is interestingly close to the observed velocities $(0.27\,c)$ of the jets in SS433 (Margon, 1984) and the jet in Cygnus X-3 ($0.16 < v/c < .31$; Molnar, Reid and Hughes, 1988). However, this may be a numerical coincidence and more work needs to be done on this point to ascertain whether this critical velocity is significant or not.

Since the morphology of class I radio sources suggests that the initial Mach number is close to unity, then this limit suggests a value of $\beta \sim 0.3$ near the bases of all class I jets (that is just beyond the so-called "gap"). This value is only an indicative value; the actual velocity may well be higher since turbulent flow will probably be initiated when the Mach number is slightly higher than unity. If the critical Mach number is 2 for example, then $\beta = 0.8$. However, even such a value of $\beta$ can imply significant deceleration.

# 4   Related Physics: Knots, Beaming and Sidedness

The question of jet-sidedness in radio galaxies is naturally related to considerations of relativistic motion. Moreover, knots in VLBI jets provide an as yet unclear diagnostic of the underlying flow. In this section some of the relevant physics is summarized and a possible reconciliation of the subluminal motion observed in a number of FRI sources with relativistic motion is given.

## 4.1   Beaming and Apparent Velocities

As emphasized by Lind and Blandford (1985) the knots in VLBI jets are naturally interpreted as shocks. Subject to some uncertainty as to whether the knot velocities represent a pattern speed or are truly indicative of the jet speed, superluminal motion of these knots has been taken as strong evidence for jet velocities at a substantial fraction of the speed of light. On the question of jet speed versus pattern speed, Hughes, Aller and Aller (1989a,b) have produced a successful model of the flux variations of BL Lac involving weak *reverse* shocks advected at approximately the jet speed. On the other hand the existence of *subluminal* knot velocities in some FRI sources (e.g. M87: $\beta_K = 0.3$ (Reid et al., 1989); Centaurus A: $\beta_K = 0.26$ (Meier et al., 1993) ; NGC 315: $\beta_K < 0.5$) raises the possibility that the jets in these sources have lower initial lower Lorentz factors. Spencer and Akujor (1992) have also had a higher success rate in finding counter jets in lower luminosity sources suggesting lower jet velocities.

The last two points suggest that a reassessment of the beaming and knot motions for low luminosity sources is required. Before embarking on this however, let me summarize, in a slightly different way than usual, the simplest relationship between beaming and apparent jet velocities.

The relations describing the surface brightness ratios of equal and oppositely directed jets and the apparent transverse velocity are well known. For the sake of completeness they are given here:

$$R = \left[\frac{1 + \beta \cos\theta}{1 - \beta \cos\theta}\right]^{2+\alpha} \quad (36)$$

$$\beta_{\mathrm{app}} = \frac{\beta \sin\theta}{1 - \beta \cos\theta} \quad (37)$$

($R$ is the surface brightness ratio, $\alpha$ is the spectral index, $\beta_{\mathrm{app}}$ is the apparent jet $\beta$ and $\theta$ is the angle to the line of sight). Powers other than $2 + \alpha$ are possible in equation (36) according to the alignment of the magnetic field. However, for the sake of simplicity and uniformity I shall restrict myself to the tangled magnetic field approximation (implying isotropic emissivity) incorporated into the drivation of equation 36. The above relations are plotted in figure 3 in the form of $R$ versus $\beta_{\mathrm{app}}$ with increments of 10° in $\theta$ marked out along each curve. The spectral index assumed is $\alpha = 0.6$. The point of this is to illustrate the restrictions that the simplest interpretation of brightness ratios and apparent velocities places on low luminosity sources when *subluminal* velocities are observed. For example in NGC 315, Venturi et al.(1993) find $R > 50$ and $\beta_{\mathrm{app}} < 0.5c$. If one assumes that the knots are moving with the jet velocity, then this implies that the jet is moving at an angle to the line of sight which is prohibitively small given the already large source size. As indicated above similar problems apply for M87 and Centaurus A.

## 4.2 Oblique Relativistic Shocks

Since the observations of subluminal velocities and the statistics of Spencer and Akujor (1992) point to slower jets in FRI sources, it is important to look at the relationship between jet speed and shock speed for jets with Lorentz factors $\lesssim 4$ (the value adopted by Hughes, Aller and Aller (1989b)). Since shocks are in general oblique I consider the following relativistic Rankine-Hugoniot relations for a shock with the structure shown in figure 4. The x axis is normal to the shock and the y-axis is in the plane of the shock. The state of gas upstream of the shock is denoted by a 1 subscript, that following by a 2. Presumably, particle acceleration and field amplification occurs at these shocks so that the emission is dominated by the post-shock region. [1]. The conservation laws for energy and momentum are (e.g. Königl, 1980):

$$w_1 \gamma_1^2 \beta_{1x} = w_2 \gamma_2^2 \beta_{2x} \tag{38}$$

$$w_1 \gamma_1^2 \beta_{1x}^2 + p_1 = w_2 \gamma_2^2 \beta_{2x}^2 + p_2 \tag{39}$$

$$w_1 \gamma_1^2 \beta_{1x} \beta_{1y} = w_2 \gamma_2^2 \beta_{2x} \beta_{2y} \tag{40}$$

Equations (38) and (40) imply that the parallel component of shock velocity is preserved: $\beta_{1y} = \beta_{2y}$. It is straightforward to solve for $\beta_{1x}$ and $\beta_{2x}$ in terms of the thermodynamic variables, by applying a Lorentz transformation to the solution for a normal shock (Landau and Lifshitz, 1987) with boost $\beta_y$ along the y axis. The result is:

$$\beta_{1x} = \gamma_y^{-1} \left[ \frac{(e_1 + p_1)(p_2 - p_1)}{(e_1 + p_2)(e_2 - e_1)} \right]^{1/2} \quad \text{and} \quad \beta_{2x} = \gamma_y^{-1} \left[ \frac{(e_2 + p_2)(p_2 - p_1)}{(e_2 + p_1)(e_2 - e_1)} \right]^{1/2} \tag{41}$$

where $\gamma_y = (1 - \beta_y^2)^{-1/2}$ is the Lorentz factor corresponding to the y-component of velocity (coming from the y-boost). For an ultrarelativistic equation of state, $p = e/3$,

$$\beta_{1x} = \gamma_y^{-1} \left[ \frac{3 p_2/p_1 + 1}{3(3 + p_2/p_1)} \right]^{1/2} \quad \text{and} \quad \beta_{2x} = \gamma_y^{-1} \left[ \frac{3 + p_2/p_1}{3(3 p_2/p_1 + 1)} \right]^{1/2} \tag{42}$$

The relationship between the angles of the pre- and post-shock fluid to the shock normal (see figure 4) are given by:

$$\tan \psi_2 = \left( \frac{\beta_{1x}}{\beta_{2x}} \right) \tan \psi_1 = \left[ \frac{3 p_2/p_1 + 1}{3 + p_2/p_1} \right]^{1/2} \tan \psi_1 \tag{43}$$

The above Rankine-Hugoniot relations can be readily Lorentz-transformed to the observer's frame. For example, the Lorentz factors of the upstream and downstream shock fluid in the observers frame are given by:

$$\gamma_{\text{fl,obs}} = \gamma_{\text{sh}} \, \gamma_{\text{fl,sh}} \left( 1 + \boldsymbol{\beta}_{\text{sh}} \cdot \boldsymbol{\beta}_{\text{fl,sh}} \right) \tag{44}$$

where $\beta_{\text{sh}}$ and $\gamma_{sh}$ refer to the shock $\beta$ and Lorentz factor in the observer's frame and the subscripts fl,obs and fl,sh refer to "fluid with respect to observer" and "fluid with respect to shock" respectively.

---

[1] These subscripts of course refer to different regions than the 1 and 2 referring to the parsec scale and kiloparsec scale in other sections of this paper

The values of the pre- and post-shock $\beta$'s and Lorentz factors in the observer's frame, are given as a function of the pressure ratio, in figure 5, for three different shock $\beta$'s (0.1, 0.3 and 0.5) and for three different angles between fluid and shock normal in the shock frame ($\psi_1 = 0$, $30°$ and $60°$). The first and last values of $\beta_{\rm sh}$ correspond to the observed value for M87 and the upper limit for NGC 315 respectively. The dashed curves correspond to the post-shock plasma and therefore determine the amount of beaming. Note that in each case, there is not much variation in the post-shock $\beta$ or $\gamma$. This is a direct consequence of relativity: For example, in the case of a normal shock, the post-shock $\beta$ in the frame of the shock varies between the weak-shock limit [2] of $\beta_{2x} = 1/\sqrt{3}$ and the strong-shock limit of $\beta_{2x} = 1/3$.

The amount of beaming implied by these curves is interesting. It is greatest for the oblique shocks, since the more oblique the shock, the larger the component of the fluid velocity that is unaffected by the shock. However, even for a normal shock, the amount of beaming can be considerable. For example, a weak shock and $\beta_{\rm sh} = 0.5$ combine to give $\gamma_{2,\rm obs} \approx 1.8$, easily enough to account for the intensity ratio of jet and counter-jet in NGC 315, given that one is now no longer constrained in the same way by the apparent velocity.

For M87, the brightness ratio of the nuclear jet is greater than 200, requiring the Lorentz factor in the emitting region to be greater than 1.6. This is possible for weak shocks when $\psi_1 = 30$ and becomes possible for stronger shocks as $\psi_1$ increases. It is interesting to note that the knot "N1" in the Reid *et al.* (1989) map made at epoch 1982.27 appears to be oblique to the flow. Indeed, Reid *et al.*show that the M87 jet does indeed show side to side oscillations. These could be indicative of an oscillating oblique shock structure.

---

[2]However, in order that some additional emission come from the post-shock region, the shock cannot be arbitrarily weak. Hughes *et al.* (1989b) found that $p_2/p_1 \approx 3$ was satisfactory for BL Lac.

# 5 Implementation of the Conservation Laws

In this section some of the mathematical and physical details relating to the implementation of the previously developed conservation laws are given. The theory is applied to two sources NGC 315 and NGC 6251 in the following sections.

## 5.1 Estimation of Jet and Lobe Pressures.

Given the pressures in the small and large scale jets and the respective jet areas it is possible to estimate the Mach number and value of $\beta$ in the large scale jet using equations (19) and (31) of § 3, assuming a quasi-steady flow from the parsec to the kiloparsec scale.

Jet pressures are notoriously difficult to estimate rigorously and since information on a low frequency spectral turnover is usually unavailable, one is forced to rely on minimum pressure or energy estimates. As we shall see using the *minimum* pressure of the parsec-scale jet gives an *upper* estimate of the Lorentz factor and in the following I adopt the minimum pressure as a fiducial value, assessing the effect of allowing the actual pressure to vary with respect to it. Generally, one assumes $p/p_{\min} > 1$. However, since the minimum pressures estimated from shocks in the core jet may be greater than the average minimum pressure in the jet, a value of $p/p_{\min} = 0.1$ is also used.

In principle, VLBI observations offer the prospect of determining the particle pressure and magnetic field independently if a turnover in the spectrum is observed. Such a turnover has been observed in the VLBI NGC 6251 jet (D.L. Jones *et al.*1986). Unfortunately, however, the jet is unresolved and values of the particle and magnetic energy density are given in terms of upper or lower limits. Moreover, plasma pressures and magnetic energy densities estimated in this way are subject to large errors because of the sensitive dependence upon angular scale (varying as $\theta^{\pm 8}$). With this caveat, Kellerman and Owen (1988) have stated that, in core sources, the ratio of particle to magnetic energy is much greater than unity. T.W. Jones (1992), on the basis of more recent data, has argued for the reverse situation. Hughes, Aller and Aller (1989a,b) in their modeling of BL-Lac have suggested that the particle pressure dominates. This question may not be settled one way or the other before sensitive, high resolution observations with the VLBA are obtained.

On the other hand minimum energy estimates of the *kpc-scale* pressure based upon equipartition may be quite reasonable. Killeen, Bicknell and Ekers (1986) showed that the minimum pressures measured along the IC 4296 jets are close to the values implied by the X-ray data. The pressures inferred from large scale radio jets in general are typical of the interstellar media pressures inferred for a number of giant ellipticals (Thomas *et al.*1986) . Moreover, the total pressure has a very shallow minimum (for $p > B^2/8\pi$) when considered as a function of, say, the ratio of particle to magnetic pressures. These assertions are contradicted to some extent by the data on NGC 1399 (Killeen, Bicknell and Ekers, 1988) since the minimum pressure is about an order of magnitude less than the confining thermal pressure. However, NGC 1399 is an extremely weak source and it is possible that the pressure in that jet is dominated by entrained thermal material. Equality between non-thermal jet pressure and thermal ISM pressure is also satisfied by more than 50% of the jets in the statistical study of Morganti *et al.* (1988). However, there are some exceptions, usually in regions of the jets close to the core. These may be over pressured regions associated with shocks and the radio resolution may be insufficient to estimate $p_\mathrm{min}$ correctly or else the extrapolation of the thermal atmosphere model close to the core may be unwarranted. Nevertheless, in the following sections I use the minimum pressure estimate for the large scale jet since any variation from this value may be incorporated into the parameter $(p_1 A_1 / p_2 A_2)$ which determines the kpc-scale velocity and Mach number (see equations (19) and (31)). This is another reason for allowing values of the pressure in the VLBI jet below its minimum value.

In the following, I also have occasion to use minimum energy estimates in the lobes when estimating the energy flux. Generally, this is probably the region of a radio source where one would most expect to find plasma in equipartition since most plasma, when left long enough, will usually generate, through turbulence, a dynamically important magnetic field. Moreover, in every jet model constructed, the ratio of magnetic energy to the particle energy increases along the jet so that the plasma delivered to the lobe may be near equipartition. Observationally, this question was addressed by the observations of Burns, Gregory and Holman (1981) who found, for a sample of radio galaxies in clusters, that the X-Ray inferred pressures of the ICM are "of the same order" as the non-thermal minimum pressure in the lobes. Burns and Balonek (1982) also came to a similar conclusion for the outer regions of two tailed radio sources. Morganti *et al.* (1988) also compared the X-ray determined ICM thermal pressures in the vicinities of the lobes of radio sources selected from the Bologna survey and compared them with the estimates of the non-thermal minimum pressure. For six out of eleven of these the ratio of the two pressures is close to unity. In some of the other sources the ratio is of the order of 10. They offered a number of possible explanations for this, perhaps the most plausible being that since these particular sources appear old and relaxed they contain a significant admixture of thermal material. If this *is* the case, then the minimum non-thermal energy in these lobes represents what has been delivered by the jet and the energy flux into the lobe can, in principle, be calculated from the minimum non-thermal energy and the age of the source.

## 5.2  Time-Averaged flow

Introducing shocks effectively makes the flow time variable. However, in a time-averaged sense, we can consider the flow as moving with velocity of $c\,\beta_1$. The high pressure region behind each shock forces it to reaccelerate to near the pre-shock velocity. Thus beaming can constrain the post-shock Lorentz factor to a certain value, but the mean Lorentz factor of the jet can be greater. Estimating the minimum pressure in the jet from the knot pressures overestimates the average pressure. Thus if we assume the ratio of the jet pressure to the minimum pressure is a certain value and we estimate the minimum pressure in the vicinity of a knot (as is frequently the case with VLBI observations) then the real ratio of pressure to minimum pressure may be a factor of a few higher.

## 5.3  The Energy Budget

The energy budget is related to estimates of pressures and magnetic fields and assumes some importance in the following in reconciling velocity estimates with the energy in relativistic particles present in the lobes. Here, I merely summarize the relevant physics (see Bicknell (1986b) and Eilek and Shore (1989) for detailed expositions). The energy $E_L$ of a lobe is given

by
$$\frac{dE_{\rm L}}{dt} = F_{\rm E} - \dot{E}_{\rm L,exp} \qquad (45)$$

where $F_{\rm E}$ is the energy flux of the jet into the lobe,

$$\dot{E}_{\rm L,exp} = \int_{\rm Lobe} \left( p + \frac{B_{\rm L}^2}{8\pi} \right) \mathbf{v}_{\rm exp} \cdot \mathbf{n}\, dS_{\rm L} \qquad (46)$$

represents the expansion losses and $E_L$ incorporates internal and magnetic energy as well as kinetic energy. Radiative losses have been neglected in this equation and this is equivalent to assuming that the lobe age is less that the radiative age. Approximately, the expansion term is given by $(p_{\rm L} + B^2/8\pi)\, v_{\rm jet} A_{\rm jet}$ where $v_{\rm jet}$ and $A_{\rm jet}$ refer to the jet velocity and area where the jet enters the lobe. Assuming (appropriately for a class I source) that the jet is transonic by the time it enters the lobe, the energy flux (incorporating Poynting Flux) is $F_{\rm E} = (4\, p_{\rm jet} + B^2/16\pi)\, v_{\rm jet} A_{\rm jet}$, and the above equation becomes, after assuming approximate equivalence of the lobe and jet parameters,

$$F_{\rm E} \approx \frac{4\,(p_{\rm L} + B_{\rm L}^2/8\pi)}{(3\,p_{\rm L} - B_{\rm L}^2/16\pi)} \left( \frac{E_{\rm L}}{t_{\rm lobe}} \right) \qquad (47)$$

where $t_{\rm lobe}$ is the age of the region of the lobe under consideration. Thus the ratio of energy flux to $E_{\rm L}\, t_{\rm lobe}^{-1}$ varies from 4/3, for a particle dominated plasma, to approximately 3 for a plasma in equipartition. Some variations in this estimate of the energy flux, of order unity, can occur if the lobe expansion or pressure are not uniform, for example. However, for order of magnitude estimates (or better) this expression is adequate.

The use of minimum energies and magnetic fields to estimate energy fluxes is much more secure than appears at first sight: When estimating an energy flux, one usually divides by an age estimated from spectral index variations and this is proportional to $B^{-3/2}$. The particle energy in a lobe, for a fixed surface brightness is proportional to $B^{-(a+1)/2}$ where $a$ is the energy index in the electron spectrum. The ratio of particle energy to age is therefore proportional to $B^{(2-a)/2}$. Since $a \sim 2$ there is not much variation in the calculated energy flux if equipartition does not hold.

### 5.4 Minimum pressure estimates for relativistic jets

The rest frame minimum pressures and energy densities depend upon the angle of the jet to the line of sight through a power of the Doppler factor. The major difference from the normal calculation is that, for a uniform jet, the surface brightness is amplified by the factor $\mathcal{D}^{2+\alpha}$ where $\mathcal{D} = \gamma_{\rm jet}^{-1}\,(1 - \beta_{\rm jet} \cos\theta)^{-1}$ where $\theta$ is the angle of the jet to the line of sight in the observer's frame. The power of the Doppler factor is geometry-dependent and the exponent adopted here represents the simplest dependence corresponding to optically thin isotropic emission from a region with a tangled magnetic field. Other dependences will not be explored here, in order to keep the number of parameters to a minimum. In addition I shall use the "tangled field" approximation on all scales as providing a useful estimate of the minimum energy parameters. Thus, treating the jet as an uniformly filled homogeneous slab, the rest frame values of the energy density of relativistic plasma and magnetic field which minimize the total rest-frame energy density are given by:

$$\epsilon_e = [2\pi\,(1+k)\,(a+1)]^{-(a+1)/(a+5)}\ K^{4/(a+5)} \qquad (48)$$
$$B = [2\pi\,(1+k)\,(a+1)K]^{2/(a+5)} \qquad (49)$$

where

$$K = I_\nu \, \mathcal{D}^{-(2+\alpha)} \, \sin\theta \, l_{\text{slab}}^{-1} \left(\frac{\nu}{2c_1}\right)^\alpha (a-2)^{-1} \left[c_9(a) \, c_5(a)\right]^{-1} E_l^{-(a-2)} \left[1 - \left(\frac{E_u}{E_l}\right)^{-(a-2)}\right]^{-1} \quad (50)$$

where the $c_i$ are the constants introduced by Pacholczyk (1970), $l_{\text{slab}}$ is the width of the slab, $E_l$ and $E_u$ are the lower and upper limits in energy of the particle spectrum, $k$ is the ratio of energy in relativistic protons to that in electrons and $a$ is the index of the particle spectrum. Energy limits on the particle spectrum are preferable to frequency limits since constraints on the spectrum due to either lack of internal Faraday rotation or the necessity for efficient scattering are most naturally expressed in terms of energy. Hence the slightly different dependence on the parameter $a$ in the above expressions compared to the more commonly used expressions, based upon frequency. Consistent with the tangled field approximation, the minimum pressure is $[(1+k)\epsilon_e + B^2/8\pi]/3$. In the following I take $k = 0$, that is there are no relativistic protons. This assumption is made self consistently throughout and conclusions are independent of the value of $k$ since all estimates of velocities, energy fluxes etc. involve ratios of energies. The only way in which this would change, would be if there were significant acceleration of relativistic protons, from the thermal pool, along the jet. Essentially, I have assumed that all of the relativistic particles emanate from the core. VLBI studies of AGN show that the jets are significantly polarized (e.g. Wardle and Roberts, 1986) so that there is an insignificant amount of thermal matter in the VLBI jet. Moreover, if the jets do not consist of electron-positron plasma, the lower cutoff in the relativistic electron Lorentz factor, $\gamma_l \gtrsim 100$. Fortunately, the minimum energy is insensitive to the actual value of the cutoff (as equation (50) shows) and a nominal value of $\gamma_l = 10$ is assumed.

A feature of the above minimum energy solution is a strong dependence upon aspect. When a relativistic jet is viewed at an inclination of 90°, the effect of "Doppler dimming" is to require larger values of rest energy density and magnetic field than would be required of a non-relativistic jet with the same surface brightness. As the inclination decreases, the required energy density and magnetic field decrease. Depending upon the Lorentz factor and the inclination angle the variation in energy density can easily be more than an order of magnitude. This variation becomes important when estimating jet parameters, particularly the energy and momentum fluxes. One consequence is that to produce given energy or momentum fluxes, the jet velocity required is greater for a jet at a small inclination than one with the identical surface brightness viewed at 90°.

A complication in the solution of the energy and momentum equations arises from the dependence of the minimum pressure upon $\beta_2$ (important for values of $\beta_2 \gtrsim 0.3$). This complication is easily dealt with by employing an iterative procedure to solve for both the minimum pressure ($p_2$) and $\beta_2$.

### 5.5 Correction of the surface brightness for resolution effects

Since, especially for VLBI observations, the resolution is often comparable to the FWHM of the jet, the surface brightness is underestimated and should be corrected. Such a correction cannot, of course, totally compensate for lack of resolution. However, the following correction is useful and is strictly valid for a jet whose transverse surface brightness profile is Gaussian and whose surface brightness is constant in the direction of the jet. Thus it should be a reasonable approximation if the surface brightness does not change appreciably on the scale of a beam. The corrected surface brightness is:

$$I_{\nu,\text{corr}} = I_{\nu,\text{app}} \left[ 1 + \left(\frac{\Phi_x}{\Phi_{\text{jet}}}\right)^2 \sin^2\chi + \left(\frac{\Phi_y}{\Phi_{\text{jet}}}\right)^2 \cos^2\chi \right]^{1/2} \qquad (51)$$

where $\Phi_x$ and $\Phi_y$ are the FWHM of the beam major-axis and minor-axis respectively, $\Phi_{\text{jet}}$ is the jet FWHM and $\chi$ is the angle between the jet and the beam major axis. Typically, for the observations used below, this correction amounts to a factor of a few. That the correction does not involve orders of magnitude in itself indicates that the minimum pressure estimates are reasonable since they depend upon the surface brightness to a small power.

## 5.6 Limitations of Method

The main limitation of this method is that the conservation laws do not accurately apply to a transonic jet far beyond the point where the jet comes into pressure equilibrium with the interstellar medium, because of the effects of buoyancy discussed earlier when dealing with the conservation of momentum. To go far beyond this point requires the relativistic and dissipative generalization of the semi-empirical approach developed in Bicknell (1986b).

# 6 Application to NGC 315

## 6.1 The Observational Data

The large scale structure of the FRI radio galaxy NGC 315 has been studied extensively by Bridle *et al.* (1979) and Willis *et al.* (1981) . The parsec-scale structure was mapped at low dynamic range by Linfield (1982) who found a one-sided core-jet. The northern jet in this galaxy has been modeled in detail by Bicknell (1986) using an adiabatic approximation and the models presented in that paper favored an initially low density ratio jet with an initial Mach number of the order of unity. The velocity at $24''$ from the nucleus was estimated to be approximately $5000 \, \mathrm{km \, s^{-1}}$. However, this estimate is based on an energy flux argument which in turn depends upon possible spectral index gradients in the northern lobe, detected by Willis *et al.* (1981). However as Willis *et al.* warned, these spectral index gradients may be instrumental so that the lobe may be younger than inferred, implying a higher velocity.

Recently, Venturi *et al.* (1993) have conducted higher dynamic range VLBI observations of the core accompanied by further VLA observations of the region near the core, confirming Linfield's observations that the parsec-scale jet is one sided but also obtaining better constraints on the sidedness ratio. This aspect of the data and the upper limit on the transverse velocities of the knots $\beta_K < 0.5$ have already been discussed in the previous section.

In this section the Venturi *et al.* data are used in conjunction with the energy and momentum equations developed in § 3 to estimate the velocity and other parameters on the large scale in terms of the parsec-scale parameters. The two main candidates for interpretation of the VLBI jet structure are:

1. The pc and kpc sidedness is intrinsic and both the VLBI and large scale jets are viewed at large angles ($\sim 90°$) to the line of sight.

2. The pc and kpc-scale sidedness is attributable to Doppler beaming and both jets are viewed at modest angles (say $\gtrsim 30°$ to the line of sight).

As a number of people have remarked when considering NGC 315, the projected source size is already quite large so that a factor of 2 increase in size is probably the maximum that can be tolerated. Thus, an inclination of $30°$ to the line of sight, is considered here as the lower limit of what seems to be reasonable and in this section, the data for NGC 315 are analyzed for two inclination angles, $\theta = 90°$ and $\theta = 30°$.

Venturi *et al.* (1993) measured the angular size of the VLBI jet at distances of 7, 11 and 27 mas from the core and found a constant full-width-half-maximum $\approx 1.5$ mas. Since this measurement is at the limit of resolution of the VLBI observations I shall use the value at 27 mas only since the jet is likely to be widest the furthest its distance from the core. In the following, I use data derived from their 1600 MHz VLBI image which are summarized in table 1. The parameters of the large scale jet are derived from their 5 and 8 GHz VLA maps and are also given in table 1. The kpc scale jet parameters are derived at angular distances 17, 27 and 62 arcseconds from the core (respectively 5.6, 8.7 and 19 kpc for $H_0 = 75 \, \mathrm{km \, s^{-1} \, Mpc^{-1}}$). The closest distance corresponds to where the jet is just starting to rapidly expand and where one expects the Mach number to near unity. The furthest distance from the core is probably too large for this conservation method to give an accurate answer, although the following results are probably indicative. Because of the effects of buoyancy, the Mach number is probably larger and the velocity lower (to conserve the energy flux) than derived here.

## 6.2 The Kpc-Scale Mach Number

Figure 6 shows the Mach number derived for the large scale jet as a function of the Lorentz factor of the parsec scale jet and for four different values of the ratio of the total to minimum pressure of the VLBI jet (remembering that is the parameter $(p_1 A_1)/(p_2 A_2)$ which determines these solutions). The curves are for two different positions along the jet ($\Theta = 17''$ and $\Theta = 27''$) and for inclinations of $90°$ and $30°$.

When $p_1/p_{\min} \approx 1$, the jet Mach number at the two different distances from the core, is of the order of unity for Lorentz factors approximately in the range $2 \lesssim \gamma_1 \lesssim 4$. As the value of $p_1/p_{\min}$ increases, the Lorentz factors for which kpc-scale transonic flow is possible, decreases, reflecting the fact that the same momentum flux can be maintained for a lower velocity as the pressure increases. Nevertheless, for the highest value of the pressure considered ($p/p_{min} = 100$) the large scale jet Mach number is unity for $\gamma_1 \approx 1.1$, that is $\beta_1 \approx 0.4$. Thus, unless the jet pressure is much higher than this, the small scale jet is *at least* mildly relativistic. On the other hand a kpc-scale Mach number $\sim 1$ is difficult to reconcile with a highly relativistic high pressure pc-scale jet, viewed at right angles. It should also be remarked that when $p/p_{\min} \sim 0.1$ transonic flow is possible for quite a wide range of Lorentz factors.

The Mach number solutions obtained for a jet at $30°$ to the line of sight stand in marked contrast to the ones just described. Transonic large scale flow is possible for a wider range of Lorentz factors and VLBI jet pressures. The reason for this of course is that the inferred rest-frame minimum energy and pressure are lower, because of beaming, so that the same energy and momentum flux require a higher velocity.

## 6.3 The Kpc-Scale Velocity

The kpc-scale velocity shows behavior consistent with the Mach number solutions just described and the relationship between Mach number and $\beta$ determined in § 3. That is, mildly relativistic velocities ($\beta \sim 0.5$) are produced for modest Lorentz factors and/or modest values of $p/p_{\min}$. The solutions are shown in figure 7. Again for $\theta = 30°$ the allowable values of $p/p_{\min}$ for mildly relativistic flow are higher for a given Lorentz factor.

## 6.4 The Kpc-Scale Value of the Density Ratio

As shown in the previous section the value of the jet density ratio can also be estimated from the solution of the jet conservation equations. The value of $\mathcal{R} = \eta/\eta_{\mathrm{crit}}$ (where as previously, $\eta$ is the jet density ratio and $\eta_{\mathrm{crit}}$ is defined by equation (21)) is shown in figure 8. The obvious point about this plot is that for parsec scale Lorentz factors which are consistent with subrelativistic flow on the large scale, the large scale values of $\eta/\eta_{\mathrm{crit}} \gtrsim 10$. Furthermore, very high values of this ratio (required by $\eta \sim 1$) demand sub-relativistic initial velocities. Since the initial velocities seem to be at least mildly relativistic, the kpc-scale value of the jet density ratio would be in the range $6 \times 10^{-5} - 6 \times 10^{-4}$, consistent with Bicknell (1986) where it was claimed that the kpc-scale jet in NGC 315 is very light near the base. Note that the value of $\eta$ increases with distance from the core, consistent with entrainment.

As the parameter $p/p_{\min}$ increases, the value of $\eta_2$ decreases, for a fixed pc-scale Lorentz factor, $\gamma_1$. This behavior is consistent with the increase of both $\beta_2$ and $\mathcal{M}_2$ with $\gamma_1$.

The major difference for the jet inclined at 30° is that the values of $\eta_2$ are higher, reflecting the lower inferred velocities at this inclination.

## 6.5 The Energy Flux

The energy flux
$$F_E = 4\,\gamma_1^2\,\beta_1\,p_1\,c\,A_1 \tag{52}$$
derived for the parsec scale jet is an important constraint since this is conserved to a very good approximation not just on the pc to kpc scale, but all along the jet. Thus the energy flux of the core jet is equal to the energy flux into the lobe and it is important to ascertain whether a relativistic core jet is consistent with lobe-based estimates of the energy flux. For the reasons outlined in § 5 minimum energy estimates of energy and magnetic field are used for the latter.

In figure 9 the energy flux is plotted against the Lorentz factor of the VLBI jet for the same values of $p/p_{\min}$ as used previously and for the two different inclinations. If the inclination of the jet is 90° and the Lorentz factor is $\gtrsim 2$ then the energy flux is at least $10^{43}\,\mathrm{ergs\,s^{-1}}$ and increases rapidly with increasing $p/p_{\min}$. For $\theta = 30°$, the energy flux curves move down by an order of magnitude so that, for example, $\gamma_1 = 3$ and $p/p_{\min} = 10$ corresponds to $10^{43}\,\mathrm{ergs\,s^{-1}}$ and $\gamma_1 = 4$ and $p/p_{\min} = 10$ gives $4 \times 10^{43}\,\mathrm{ergs\,s^{-1}}$. This again reflects the lower value of $p_{\min}$ due to beaming.

Let us now examine whether it is feasible that these energy fluxes are consistent with the energy flux into the northern lobe. The region of the northern lobe which I consider is the region approximately $240''$ in length to the west of the northern jet and approximately perpendicular to it. This is of reasonably high surface brightness and the Willis *et al.* (1981) map can be used to obtain minimum energy and pressure estimates. The minimum energy for this region is approximately $6 \times 10^{57}$ ergs. Thus, allowing a nominal factor of two for expansion work, the energy flux into this region is approximately $4 \times 10^{42}\,t_8^{-1}\,\mathrm{ergs\,s^{-1}}$ where $t_8$ is the age in units of $10^8$ yr. This estimate at least puts the energy flux in the vicinity of $10^{42}\,\mathrm{ergs\,s^{-1}}$ since estimates of a few $\times 10^8$ yr are common for FRI lobes, although one also has to note that this region is probably a small part of the extended emission from NGC 315. There are other fainter features visible in the Willis *et al.* map and $10^8$ yr may be an overestimate.

Willis *et al.* report some spectral index variations from this region of the source but caution that they may be instrumental. Thus, another method for estimating $t_8$ is required: Let us assume that the flow of jet material into the lobe is transonic and subrelativistic at this point (reasonable, in view of the physics which is thought to apply in class I jets and supported by the existing models). Furthermore, consider the expression (equation (30)) for the mass flux in this limit:

$$F_\mathrm{M} = \frac{1}{4}\left(\frac{\mu m_p}{kT}\right) F_E$$

$$= 3\,M_\odot\,\text{y}^{-1}\,\eta\left(\frac{F_\text{E}}{10^{42}\,\text{ergs s}^{-1}}\right) \qquad (53)$$

Clearly, the jet density ratio cannot be near unity near the end of the jet, for otherwise the mass flux would be unreasonable. In fact, this relation seems to point to a density ratio $\lesssim 0.1$ which would imply that the mass flux does not exceed the stellar mass-loss rate in the galaxy. Hence, the fluid velocity at the end of the jet is $\approx 1,400\,\text{km s}^{-1}\,(\eta/0.1)^{-1/2}\,\mathcal{M}\,(T_{\text{ISM},7})^{1/2}$. The morphology of the western side of NGC 315 suggests that the jet plasma continues to flow to the south west and for the region under consideration, let us assume that this velocity is approximately constant. The (projected) size of the region of the lobe is approximately 150 kpc, giving an age $t_\text{lobe} \approx 1.1 \times 10^8\,\text{yr}\,(\eta/0.1)^{1/2}\,\mathcal{M}^{-1}\,T_{\text{ISM},7}^{-1/2}$ and the energy flux, therefore, is approximately $3.5 \times 10^{42}\,\text{ergs s}^{-1}\,(\eta/0.1)^{-1/2}\,\mathcal{M}\,T_{\text{ISM},7}^{1/2}$. There are various factors that could push this value up towards $10^{43}\,\text{ergs s}^{-1}$: a lower density ratio and an interstellar medium temperature of a few$\times 10^7\,K$. However, even a value of $4 \times 10^{42}\,\text{ergs s}^{-1}$ puts the $\theta = 30°$ jet in a range of Lorentz factors and core pressures that are consistent with a relativistic core jet and mildly relativistic flow on the kpc scale.

## 6.6 Beaming

As shown in the previous section, Doppler beaming on the parsec scale could be reconciled with subluminal motion if the knots in the NGC 315 jet are reverse shocks. On the kiloparsec scale the jets in NGC 315 are also asymmetric and it is of interest to see if this can also be due to relativistic beaming.

The Venturi et al.(1993) VLA maps of the region near the core show that when the ratio of the main jet to counter jet surface brightnesses at $\Theta = 27''$ is approximately 20 implying a kpc scale $\beta \cos\theta \approx 0.5$ at this point. If the jet is at 45° to the line of sight $\beta \approx 0.7$; at $\theta = 30°$, $\beta \approx 0.6$. These values are of interest for two reasons: Firstly, inspection of the $\theta = 30°$ model (figure 7) shows that $\beta = 0.6$ at $\Theta = 17$ implies a pc-scale Lorentz factor of approximately 3.2 for $p/p_\text{min} = 10$, corresponding to an energy flux $\approx 10^{43}\,\text{ergs s}^{-1}$. If the jet were inclined at a greater angle, a smaller Lorentz factor or value of $p/p_\text{min}$ would be required, for a similar value of $\beta$ and the same energy flux. The implication for the energy flux can be derived independently from the kpc-scale data since the energy flux is conserved. Given the assumption of mildly relativistic flow in the rapidly expanding section, the energy flux is dominated by the term $4\,p\gamma^2\,\beta c\,A$ since $\mathcal{R} \sim 1 << \gamma/(\gamma-1)$ and the minimum value of the energy flux can be estimated by substituting $p_min$. For example, at $\Theta = 17$, assuming $\theta = 30°$, and $\beta \approx 0.6$ implies $p_\text{min} = 1.4 \times 10^{-11}\,\text{dy cm}^{-2}$ and $F_E \approx 1.2 \times 10^{43}\,\text{ergs s}^{-1}$. As stated earlier, there is good evidence for the proposition that large scale jets are near the minimum pressure, so that such an estimate is reasonable.

The second point about the beaming-based estimate of $\beta$ in the large scale jet is that in the rapidly spreading region of the northern jet which starts at approximately 15″ from the core, the Mach number is arguably between 1 and 2 since that is the region of Mach numbers for which jets are most turbulent. According to the previously derived relationship between jet velocity and Mach number (see figure 2), this corresponds to $0.3 < \beta < 0.7$. This result is independent of pressure estimates and such consistency between values of the jet $\beta$ inferred from the surface brightness constraint and dynamical arguments which have been made as model-independent as possible, strongly supports the notion that the jet surface brightness asymmetry in NGC 315 is due to beaming.

The subsequent behavior of the surface brightness asymmetry is also of interest. Figure 10 (from data in Willis *et al.*(1981)) shows the integrated surface brightness of the main jet and counter jet. By approximately $100''$ these are within a factor of 2 of each other, implying $\beta \lesssim 0.17$. This is consistent with the models presented in Bicknell (1986b) which all show a decline in velocity (more or less independent of jet density ratio) by a factor of approximately 3.5 between $\Theta = 24''$ and $\Theta = 100''$. In fact this is also, to a large extent, model independent, since this is the decline in jet velocity required to model an adiabatic surface brightness decline. Other features in this plot, for example, the difference in surface brightnesses at the ends of the jets are probably environment dependent.

# 7 Application to NGC 6251

In this section data from various sources on NGC 6251 is analyzed in a similar way to that on NGC 315.

## 7.1 The Observational Data

The first radio image of this class I radio galaxy was made by Waggett, Warner and Baldwin (1977) and this was followed by more detailed observations of the northern jet by Saunders et al. (1981). The source was subsequently observed at high resolution with the VLA by Perley, Bridle and Willis (1984) and was studied at VLBI resolution by D.L. Jones et al. (1986). The VLBI and VLA jet data used in the following are summarized in table 2.

One difference between NGC 6251 and the NGC 315 VLBI data is that Jones et al. actually detected a low frequency turnover in the spectrum of the jet. However, the jet is unresolved ($\Phi_1 < 0.5$ mas) so that they could only constrain the magnetic field and particle energy to satisfy $B^2/8\pi < 0.2 \,\mathrm{ergs\,cm^{-3}}$ and $\epsilon_e > 6 \times 10^{-7} \,\mathrm{ergs\,cm^{-3}}$ respectively. These constraints (for a $\gamma = 1$ jet) are unlikely to be very tight because of the sensitivity of the magnetic field and particle energy density to the width of the jet. Thus in this section, I adopt the same approach of varying the minimum energy estimate for the magnetic field and pressure. Because of the uncertainty in the jet width, I also use two values for this parameter $\Phi_1 = 0.5$ mas and $\Phi_1 = 0.25$ mas.

Another important difference between NGC 315 and NGC 6251, on the large scale, is that the jet morphology beyond $\Theta \approx 18''$ possibly does not indicate a transonic Mach number. The overall rate of expansion of the jet is lower (indicative of higher Mach number flow) and there are knots of emission indicating local dissipation, presumably through shocks. Thus the NGC 6251 jet appears not to be the "gentle" turbulent flow envisaged for lower-powered class I jets in which dissipation may occur through Fermi acceleration and weak shocks (Bicknell and Melrose, 1982). However, it is possible that in the region where the jet does rapidly expand (the region up to $18''$ from the core) the jet Mach number is near unity but that beyond this point, the Mach number increases because of a steep pressure gradient. (In NGC 6251, the jet minimum pressure decreases approximately as $R^{-3}$ where $R$ is the distance from the core.) Since for NGC 6251, the Mach number is arguably of the order of a few, the most important constraint on the jet Lorentz factor is the energy flux. Nevertheless, it is of interest to examine the other parameters, in particular the velocity, to examine under what conditions it is possible to decelerate a relativistic core jet to a sub-relativistic or mildly relativistic flow some kiloparsecs from the core. In so doing, it is worthwhile keeping in mind the constraints on the velocity and inclination implied by a beaming interpretation of both the VLBI and VLA data. The lower limit on the VLBI counterjet surface brightness implies $\beta \cos\theta > 0.70$, (that is $\gamma > 1.4$) and an inclination less than $45°$ (D.L. Jones et al. (1986)). The observed faint counterjet is 65 times fainter than the main jet $30''$ from the core, implying $\beta \cos\theta = 0.66$ ($\gamma > 1.3$). For $\theta = 30°$ these limits correspond to ($\beta > 0.81$, $\gamma > 1.7$) and ($\beta = 0.76$, $\gamma = 1.5$) respectively.

## 7.2 The Kpc-Scale Mach Number

Figure 11 shows the large scale Mach number as a function of the pc-scale Lorentz factor. The trends are the same as for NGC 315 except that for a given Lorentz factor, the Mach number is generally higher. In the $\theta = 90°$ jet, Mach numbers around 1 or 2 are feasible for Lorentz factors $\sim 2-3$ and $p/p_{\min} = 1.0$ When $p/p_{\min} = 10$ the Lorentz factors for which this is possible decrease to $\sim 1.1$. The same inclination effect is present: the Mach numbers for a given Lorentz factor are lower at the 30° inclination. On the other hand, the Mach numbers implied by Lorentz factors of 2-3 are perhaps not that unreasonable, given the morphology discussed above.

## 7.3 The Kpc-Scale Velocity

Figure 12 shows the kpc-scale value of $\beta$ as a function of the pc-scale Lorentz factor. All of these plots show that for pc-scale Lorentz factors $\sim 2-3$, the velocity at $\Theta = 10''$ is at least mildly relativistic but that substantial deceleration takes place by $\Theta = 20''$. The corresponding Lorentz factors increase if the jet diameter $\Phi_1 = 0.25$ mas.

The inferred Mach numbers and velocities for a given Lorentz factor are greater in NGC 6251 than they are in NGC 315. This is due to the larger minimum pressure and smaller width of the NGC 6251 VLBI jet and reinforces one's subjective impression, based upon morphology that NGC 6251 is a faster jet.

Further comments are made upon the Mach number and velocity in the subsection below relating to the energy flux.

## 7.4 The Density Ratio

The density ratio (figure 13) shows behavior consistent with the velocity and Mach number. For all but the lowest Lorentz factors, the actual density ratio would be quite low ($\sim 10^{-5}$).

## 7.5 The Energy Flux

As with NGC 315, the energy flux places an important constraint on the initial Lorentz factor of the NGC 6251 jet. Figure 14 shows the energy flux as a function of the initial Lorentz factor $\gamma_1$ and for the same values of the ratio of the jet pressure to the minimum pressure. Saunders et al.(1981) estimated $E_{\min} \approx 2.0 \times 10^{58}$ ergs (their result scaled from $H_0 = 50$ to $H_0 = 75$). Saunders et al.also quoted significant spectral steepening by 10.7 GHz. Combined with the minimum energy magnetic field $\approx 3.2 \mu G$ and assuming no extra energy redistribution from particle acceleration, this implies an age of the north western lobe $\sim$ the synchrotron plus inverse Compton cooling time,

$$\begin{aligned} t_{\text{cool}} &\approx 5.2 \times 10^9 \text{ yr } \frac{B_{-6}^{-3/2}}{1 + \left(\frac{B_{IC}}{B}\right)^2} \nu_9^{-1/2} \\ &\approx 1.3 \times 10^8 \text{ yr} \end{aligned} \quad (54)$$

Thus, allowing a nominal factor of 2 for expansion work, the energy flux is approximately $10^{43}$ ergs s$^{-1}$.

Figure 14 obviously shows that it is easier to sustain an argument for a relativistic core jet for a range of $p/p_{\mathrm{min}}$ if the inclination is closer to 30°. As with NGC 315, the energy flux curves are displaced downwards for the same Lorentz factor and $p/p_{\mathrm{min}}$. For $\gamma_1 > 1.7$ there is a range of $p/p_{\mathrm{min}}$ between 1 and 10, for which the global energy budget is satisfied. The range of Lorentz factors is higher when the core jet diameter is 0.25 mas rather than 0.5 mas.

When the inclination is 90°, for most Lorentz factors greater than 2, and for either value of the assumed jet diameter ($\Phi_1 = 1.25$ or 0.5 mas) the value of $p/p_{\mathrm{min}}$ is required to be less than unity in order to produce an energy flux consistent with beaming of both the pc-scale and kpc-scale jets. Thus, consistency of the beaming explanation and the source energetics require an inclined jet. An inclination of 30° provides a comfortable range of pressures and Lorentz factors consistent with the energy budget but clearly other inclinations not too much greater will work as well.

As with NGC 315, the kpc-scale estimates of minimum pressure support the case for at least a mildly relativistic jet on the kpc scale. For example, at $\Theta = 30°$, the minimum pressure in a $\beta = 0.76$, $\theta = 30°$ jet, is $1.0 \times 10^{-11}$ dy cm$^{-2}$ and the enthalpy part of the energy flux ($4\gamma^2 \beta p c A$) is $1.4 \times 10^{43}$ ergs s$^{-1}$, as required.

# 8 Discussion

The main aim of this paper was to ascertain whether the milliarcsecond and arcsecond data on two radio galaxies, NGC 315 and NGC 6251 are consistent with the deceleration of a relativistic core jet to the extent that it is mildly relativistic or subrelativistic on the kiloparsec scale and whether these data support the notion that FRI radio galaxies are the parent population of BL-Lac objects. A reasonable assumption underlying this approach is that the rapid spreading rate of FRI jets on the kiloparsec scale (Bridle, 1984) indicates a transonic Mach number, although this has yet to be demonstrated for the extremely low density relativistic flows envisaged here. The mechanism envisaged for the deceleration is entrainment initiated when the jet comes into pressure equilibrium with the interstellar medium. It should also be noted that Phinney (1983) showed, that a jet with kinetic luminosity $\lesssim 10^{42}$ergs s$^{-1}$ (but possibly an order of magnitude larger, depending upon the model parameters) will be decelerated by the injection of mass into the flow due to mass-loss from stars along the path of the jet. Thus Phinney's mechanism may also be at least partially responsible for establishing the transonic flows considered here. The analysis of the jet conservation laws would be unchanged in this case since there would be no change to the mass-flux conservation law when matter is injected rather than entrained and the injected energy is negligible compared to the energy in the flow.

The analysis of the energy and momentum equations in § 5 shows, without reference to any particular set of data, that if a relativistic jet is decelerated to a transonic Mach number, then its velocity is necessarily in the range $0.3 - 0.7\,c$ at that point. Thus if the rapid spreading of most FRI jets is indicative of turbulent entrainment in a transonic jet, the jet velocities at the beginning of the turbulent region are necessarily of this order. Conversely, if an initially relativistic jet remains supersonic, then it must also remain relativistic. The plot of Mach number versus jet $\beta$ in § 5 quantifies Scheuer's (1983) assertion of this result.

This result alone adds considerable weight to the idea that the kpc scale bases of FRI jets are mildly relativistic flows and that the surface brightness asymmetries observed in FRI jets near their bases are the result of Doppler beaming.

The detailed analysis of the data on NGC 315 and NGC 6251 provides further support for this idea. Provided that the jet pressures are not too high compared to the minimum value, pc-scale jets with Lorentz factors of the order of 2 or greater, provide the right energy and momentum flux to be consistent with mildly relativistic flow on the large scale. The analyses for two different inclination angles, $\theta = 90°$ and $\theta = 30°$, favor the inclined jet as the one which is most consistent with the ideas being examined and this in turn is consistent with the beaming explanation for the jet to counter-jet surface brightness ratios in these sources. In every instance, this is due to the lower minimum pressure implied by Doppler beaming so that a required power and thrust is consistent with a higher velocity for a given $p/p_{\min}$.

The galaxies examined here are on the borderline in power between FRI and FRII. Galaxies of lower power are more likely to become subrelativistic closer to the core and as we have seen, supersonic class II jets must be highly relativistic. Therefore, the notion that the transition from two-sided jets to one-sided jets as one passes the FRI/II break, is due to the transition from subrelativistic to relativistic flow, is strongly supported by this analysis.

These calculations also confirm some important aspects of the modeling of the kpc-scale NGC 315 jet (Bicknell 1986b). In particular a low density ratio near the core (on the kiloparsec scale) is unavoidable. The models presented in that paper did not examine density ratios as low as those considered here, since there is effectively a low density limit, represented by the "hot jet" model presented in Bicknell (1984). However, it was clear that the models implied that the density is very low. The deceleration of the NGC 315 jet beyond its initial turn-on point is also supported by the dynamical analysis of this paper and by the convergence of the surface brightnesses of the jet and counter jet. The main feature of the previous modeling which is *not* supported by this paper is the estimate of the jet velocity. This was based upon the energy flux argument and the use of spectral index data which, as noted by Willis *et al.* (1981), may be suspect. In view of this, further spectral index mapping of the lobes of NGC 315 (perhaps at higher frequency) would be of interest.

Although the analysis presented here is generally supportive of mildly relativistic flow on the large scale in FRI jets and explanations for surface brightness asymmetries in terms of beaming, there are some caveats. The projected sizes of NGC 315 and NGC 6251 are large and beaming requires that they be larger, perhaps by up to a factor of 2. Nevertheless, this may be simply an unavoidable consequence of more compelling physics. Subluminal motion raises another problem. One way of dealing with this has been given in § 4: Reverse shocks, although advected with the flow, give a misleading indication of the flow velocity which in fact can be large enough to give significant beaming on the pc-scale, and the requisite power and thrust to support mildly relativistic flow on the kpc-scale. In the case of NGC 315 and NGC 6251, this is just possible with normal reverse shocks. It is much easier if the shocks are oblique. Thus higher resolution VLBI observations as well as the determination of proper motions in the core jet are important. The energy flux in any jet and in particular, NGC 315 and NGC 6251, is an important constraint and in order that the core jets be relativistic the required energy flux is of the order of $10^{43}$ ergs s$^{-1}$ in each case. This is feasible but it would be interesting to have independent verification. The fact that the determination of spectral index gradients by low resolution observations of class I lobes can shed light on the Lorentz factors of core jets, is perhaps worthy of reflection.

The idea that FRI's are the parent population of BL Lacs is also supported by this analysis. Whilst the knot motions in NGC 315 are subluminal and BL Lacs are usually associated with superluminal motions with $\beta_{\rm app} \sim 2 - 4$ the plausibility of relativistic flow on the parsec scale has been established in the range of extended radio luminosities associated with BL Lac objects. Nevertheless, the existence of subluminal motions in a number of FRI sources and the differences between knot velocities and the velocity of the material responsible for the beamed emission indicates that the current models used for estimating the BL Lac luminosity function from the FRI luminosity function, may require some modification. This may involve little more than acknowledging that the Lorentz factor used in the luminosity function calculations represents that of the post-shock material and that it is higher than that associated with the observed knot motions, possibly going some way to reconciling the Lorentz factors required by Padovani and Urry (1990) with the superluminal velocities summarized by Mutel (1992). A more detailed examination of BL Lac luminosity function models is beyond the scope of this paper.

It has been shown in §§ 6 and 7 that for some choices of jet parameters (principally a low $p/p_{\rm min}$) the Lorentz factors in the NGC 315 and NGC 6251 pc-scale jets could be greater than 5. However, generally one is more comfortable with Lorentz factors in the vicinity of 2 to 4. If the proper motions of knots are correlated with Lorentz factor, then it is arguable that the Lorentz factor of the NGC 315 core jet is closer to 2. For a shock velocity of $0.5\,c$, a pressure ratio across the shock of 3 (as used by Phillips *et al.*, 1989b in modeling BL Lac) implies, for a normal shock, a pre-shock Lorentz factor of approximately 2.6, and a post-shock Lorentz factor of 1.6. The former can give adequate thrust and power to the jet for a high enough pressure and the latter can give adequate beaming. These numbers are, of course, based upon upper limits; firmer estimates of proper motions and jet to counterjet surface brightness ratios would be welcome.

Observations of the core variability in each galaxy would also be useful, since this would enable one to apply the Hughes *et al.* (1989b) model to constrain the pressure variation in the knots and the angle of inclination.

The overwhelming *impression* that one gains from analyzing the data on these two galaxies and from the general results derived in § 4 is that the notions of jet sidedness due to beaming, mildly relativistic decreasing to subrelativistic kpc-scale velocities of FRI jets, the physics of the variability in BL Lac objects and the unification of BL Lac Objects and FRI Radio Galaxies all appear reasonably consistent. This impression is, of course, subject to the caveats mentioned above and it will be interesting to see the results of future VLBI observations of FRI radio galaxies, particularly as they pertain to the relative motions of knots and jets and, if possible, to the absolute value of the pressure. As I stated in the introduction, the computer programs used for this paper are available to anyone who wishes to use them to analyze the relationships between pc-scale and kpc-scale jets.

The analysis in this paper has been mainly based on the popular assumption that VLBI jets are initially free. This may not be the case if they are surrounded (for example) by a magnetically collimated (but more slowly moving) wind (Blandford, 1994) . Thus one could envisage a high Mach number jet gradually decelerating until it became transonic. Little would change in the analysis since in the high Mach number regime, the jet momentum would still be approximately conserved. The details of such a scenario will be examined in a future paper.

**Acknowledgement** This work was stimulated by the paper given by Dr. R.A. Laing at the 1992 STScI Symposium on Astrophysical Jets. I am also indebted to Dr. Laing for many two-sided and informative discussions on this topic. I am also grateful to Dr. T. Venturi and her colleagues for communication of their data on NGC 315 prior to publication.

# 9 Tables

### Table 1 - Summary of Venturi *et al.*(1993) NGC 315 Data

| $\Theta$ | $R$ | $\Phi_{obs}$ | $\Phi_{decon}$ | $\nu_{obs}$ (MHz) | $S_\nu$ (mJy/beam) | Beam | Posn. Angle (degrees) |
|---|---|---|---|---|---|---|---|
| 27 mas | 8.7 pc | - | 1.5 mas | 1600 | 1.75 | 7.5mas $\times$ 2.2mas | -9 |
| 17″ | 5.6 kpc | 3.5″ | 3.3″ | 5000 | 2.0 | 1.3″ $\times$ 1.3″ | - |
| 27″ | 8.7 kpc | 8.1″ | 7.0″ | 8400 | 6.0 | 4.2″ $\times$ 3.8″ | -64 |
| 62″ | 19.9 kpc | 16.2″ | 15.7″ | 8400 | 1.0 | 4.2″ $\times$ 3.8″ | -64 |

### Table 2 - NGC 6251 Data

| $\Theta$ | $R$ | $\Phi_{decon}$ (arcsec) | $\nu_{obs}$ (MHz) | $S_\nu$ (mJy/beam) | Beam | Reference |
|---|---|---|---|---|---|---|
| 5 mas | 2.2 pc | $\lesssim$ 0.5mas | 4989 | 22 | 1.3 mas $\times$ 1.1 mas | D.L. Jones *et al.*(1986) |
| 10.0″ | 4.3 kpc | 1.4″ | 1662 | 1.24 | 1.15″ $\times$ 1.15″ | Perley, Bridle |
| 20.0″ | 8.6 kpc | 3.0″ | 1662 | 2.15 | 1.15″ $\times$ 1.15″ | and Willis (1984) |
| 30.0″ | 12.9 kpc | 3.5″ | 1662 | 3.5 | 1.15″ $\times$ 1.15″ | ″ |

# 10 Figures

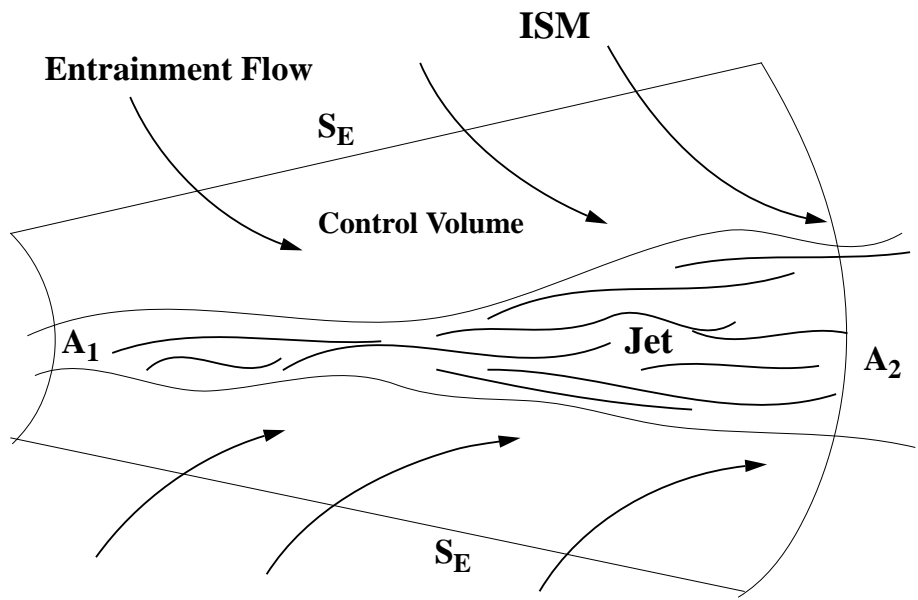

Figure 1: The control volume for the conservation laws derived in § 3. The "entrainment surface" $S_E$ is at a far enough distance from the jet that external conditions apply. The jet cross-sectional areas $A_1$ and $A_2$ are indicated.

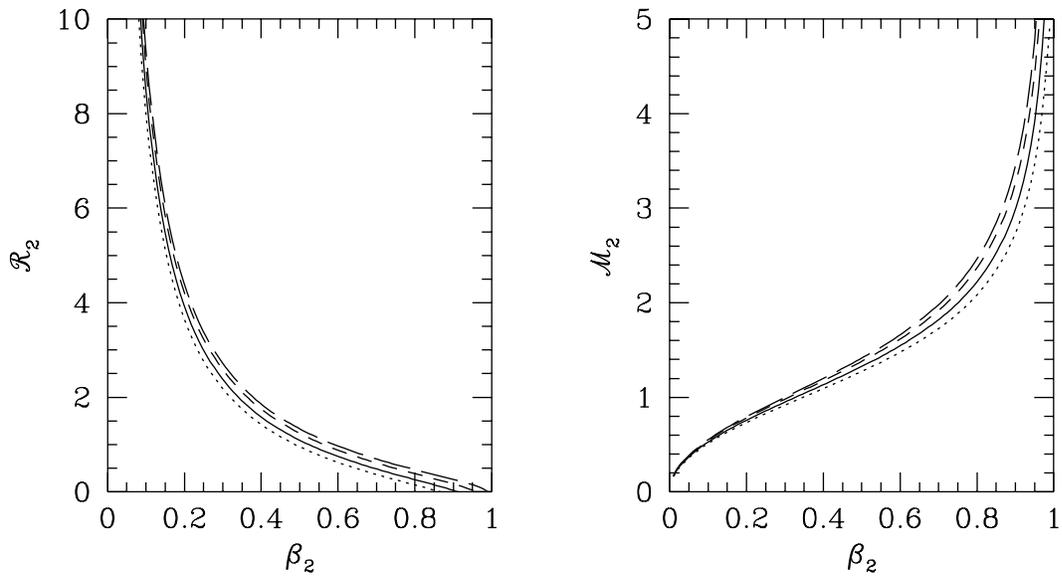

Figure 2: Left panel: The value of the ratio of rest-mass energy to enthalpy, $\mathcal{R}$, as a function of the jet $\beta$. Right panel: The Mach number as a function of the jet $\beta$. Solid curves: $\gamma_1 = 5$; dotted curves: $\gamma_1 = 2.5$; Long-dashed curve: $\gamma_1 = 1.25$; Short-dashed curve: $\gamma_1 = 1.1$.

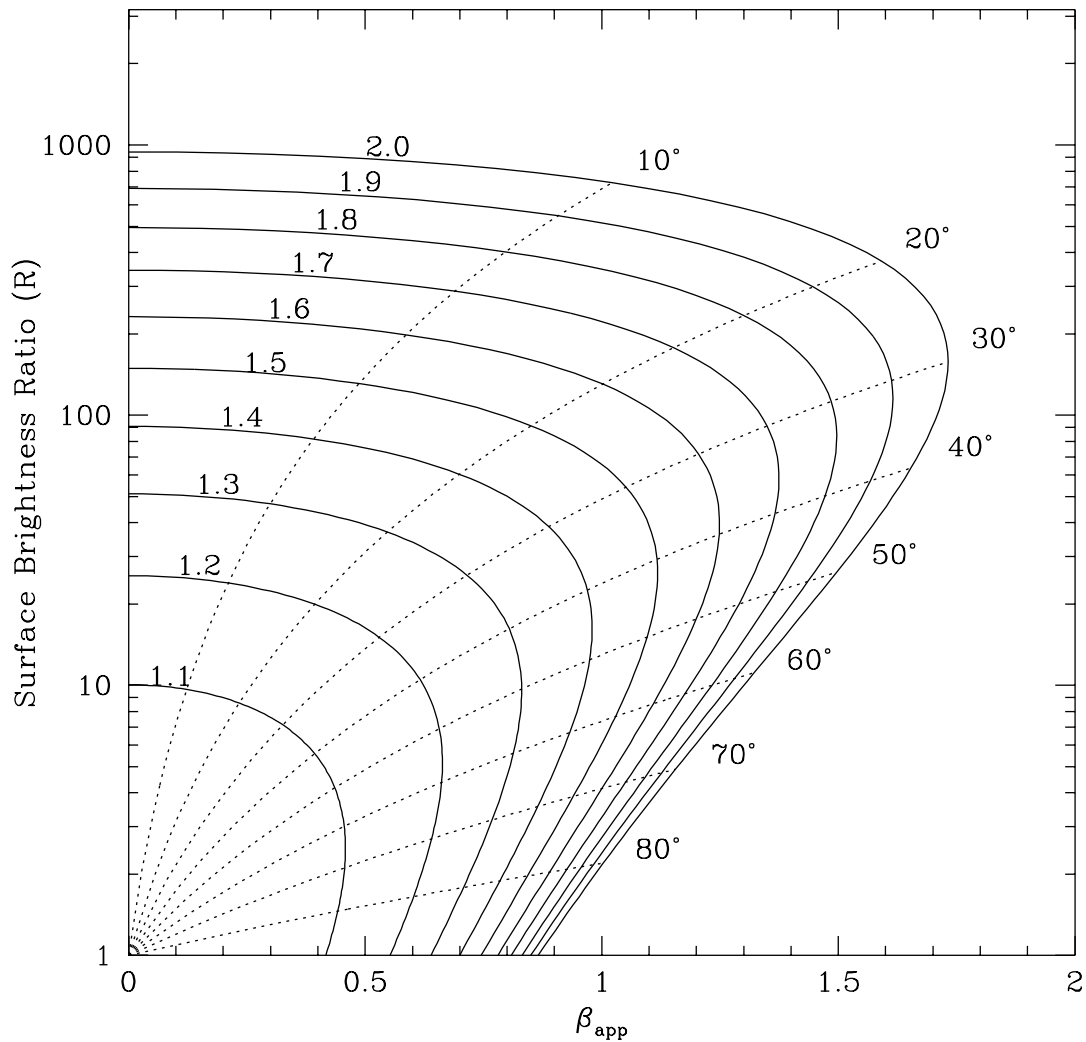

Figure 3: The brightness ratio of two oppositely directed relativistic jets as a function of the apparent $\beta$ of the jet. The solid curves correspond to the indicated Lorentz factors with the angle of inclination, $\theta$, varying between $0°$ and $90°$. The dotted curves represent the loci correpsonding to a fixed angle of inclination, but varying Lorentz factor. The different angles are marked at the extremities of the curves. Note the restricted range of $\theta$ for brightness ratios greater than 50 and apparent velocities less than 0.5 c.

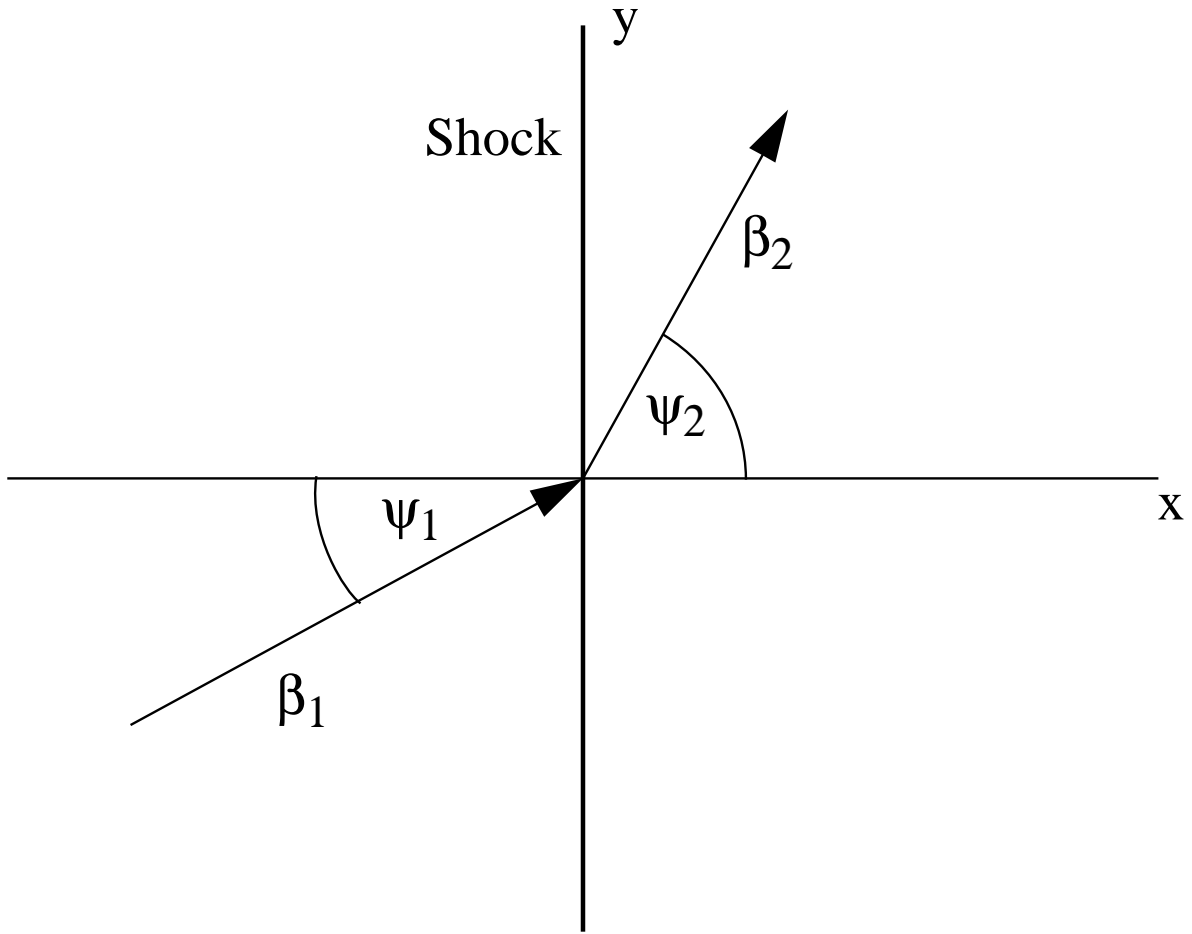

Figure 4: The structure of a relativistic oblique shock illustrating the symbols used in the text (following Königl, 1980).

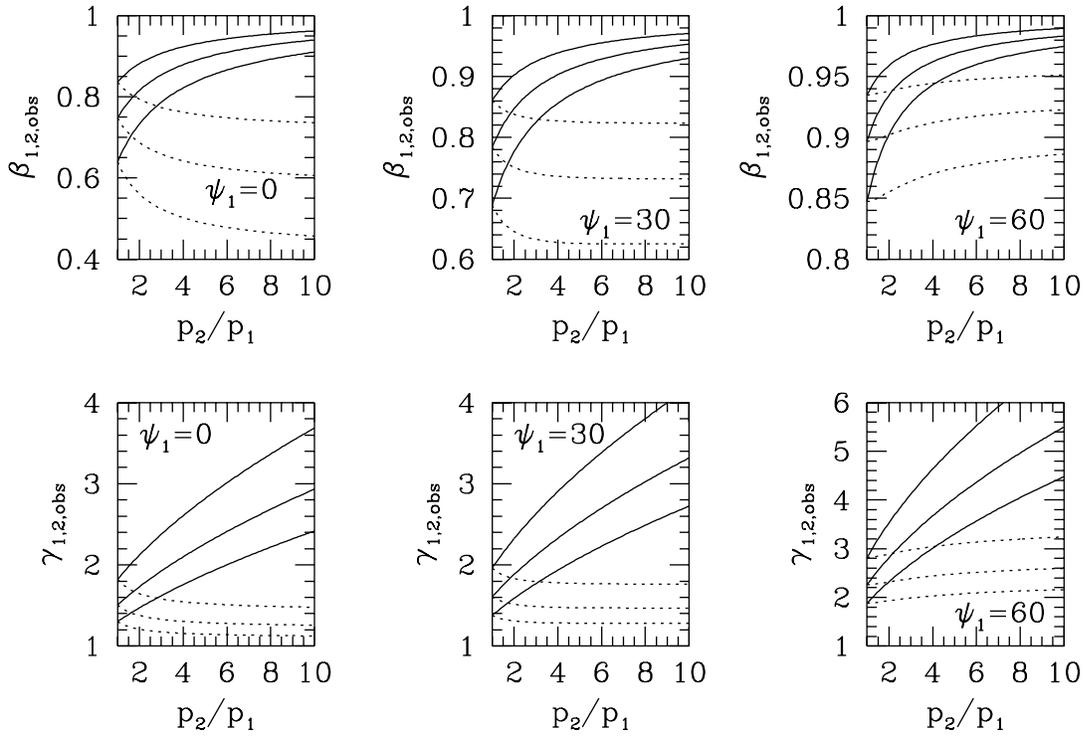

Figure 5: The upstream and downstream values of the jet $\beta$ and Lorentz factor for three different values of the observed shock velocity ($\beta_{\rm sh}$). The solid curves represent the upstream values and the dashed curves represent the downstream ones. The lowest pair of curves corresponds to $\beta_{\rm sh} = 0.1$; as the shock velocity increases, each pair of curves is translated upwards.

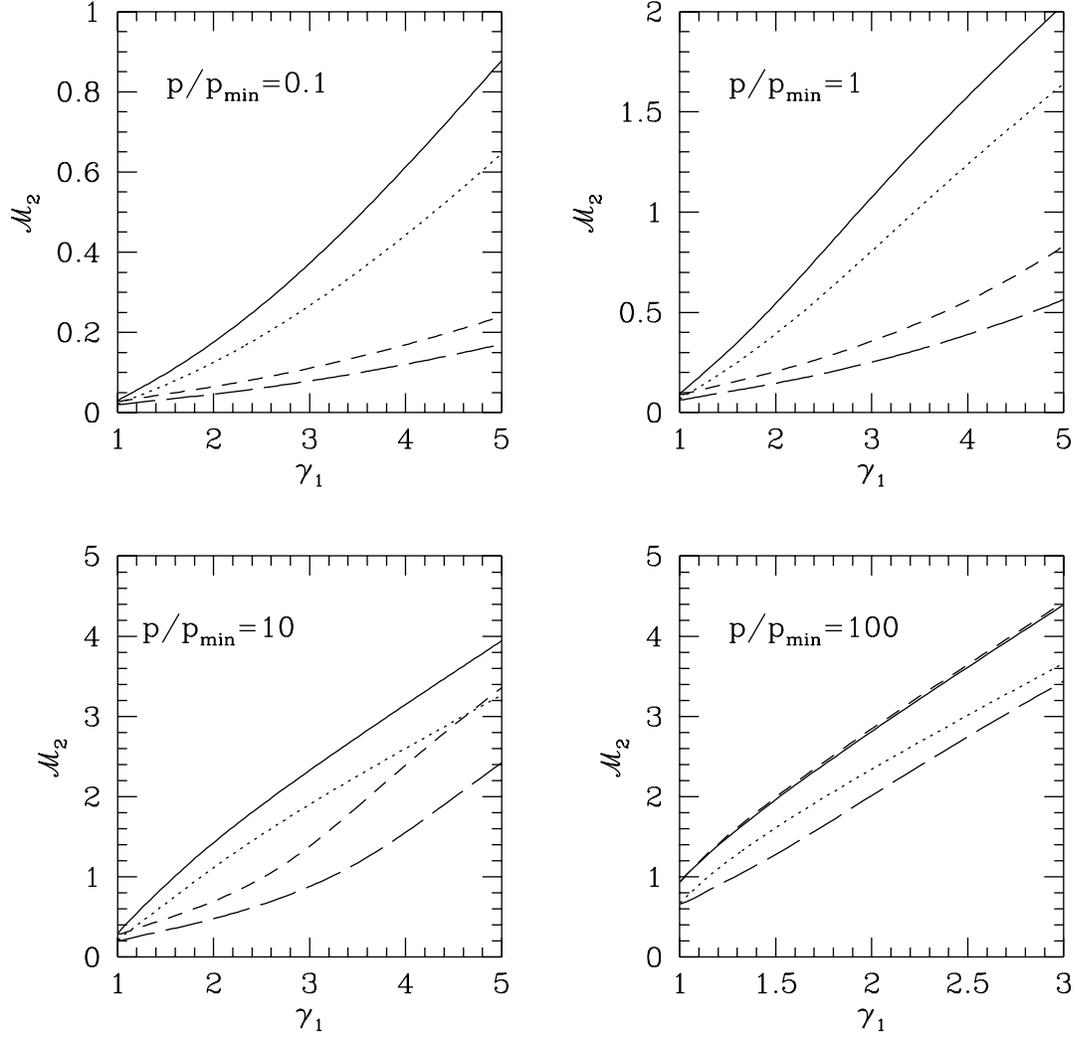

Figure 6: The Mach number at 2 different positions along the NGC 315 kpc-scale jet as a function of the pc-scale Lorentz factor for different values of $p/p_{\min}$ and jet inclination, $\theta$. Solid curve: $\Theta = 17''$, $\theta = 90°$, dotted curve: $\Theta = 27''$, $\theta = 30°$, short-dashed curve: $\Theta = 17''$, $\theta = 90°$, long-dashed curve: $\Theta = 27''$, $\theta = 30°$.

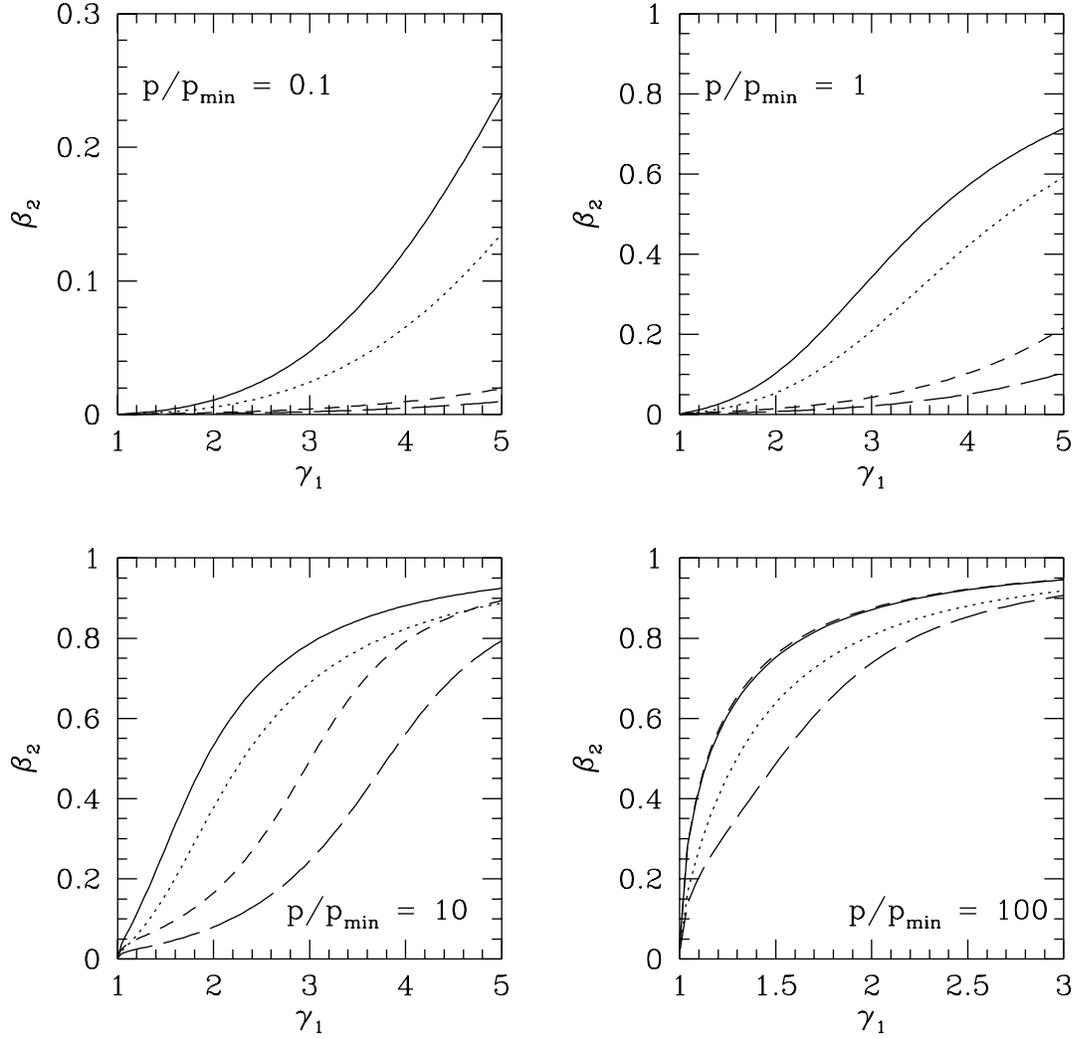

Figure 7: The value of $\beta$ at 2 different positions along the NGC 315 kpc-scale jet as a function of the pc-scale Lorentz factor for different values of $p/p_{\min}$ and jet inclination, $\theta$. Solid curve: $\Theta = 17''$, $\theta = 90°$, dotted curve: $\Theta = 27''$, $\theta = 30°$, short-dashed curve: $\Theta = 17''$, $\theta = 90°$, long-dashed curve: $\Theta = 27''$, $\theta = 30°$.

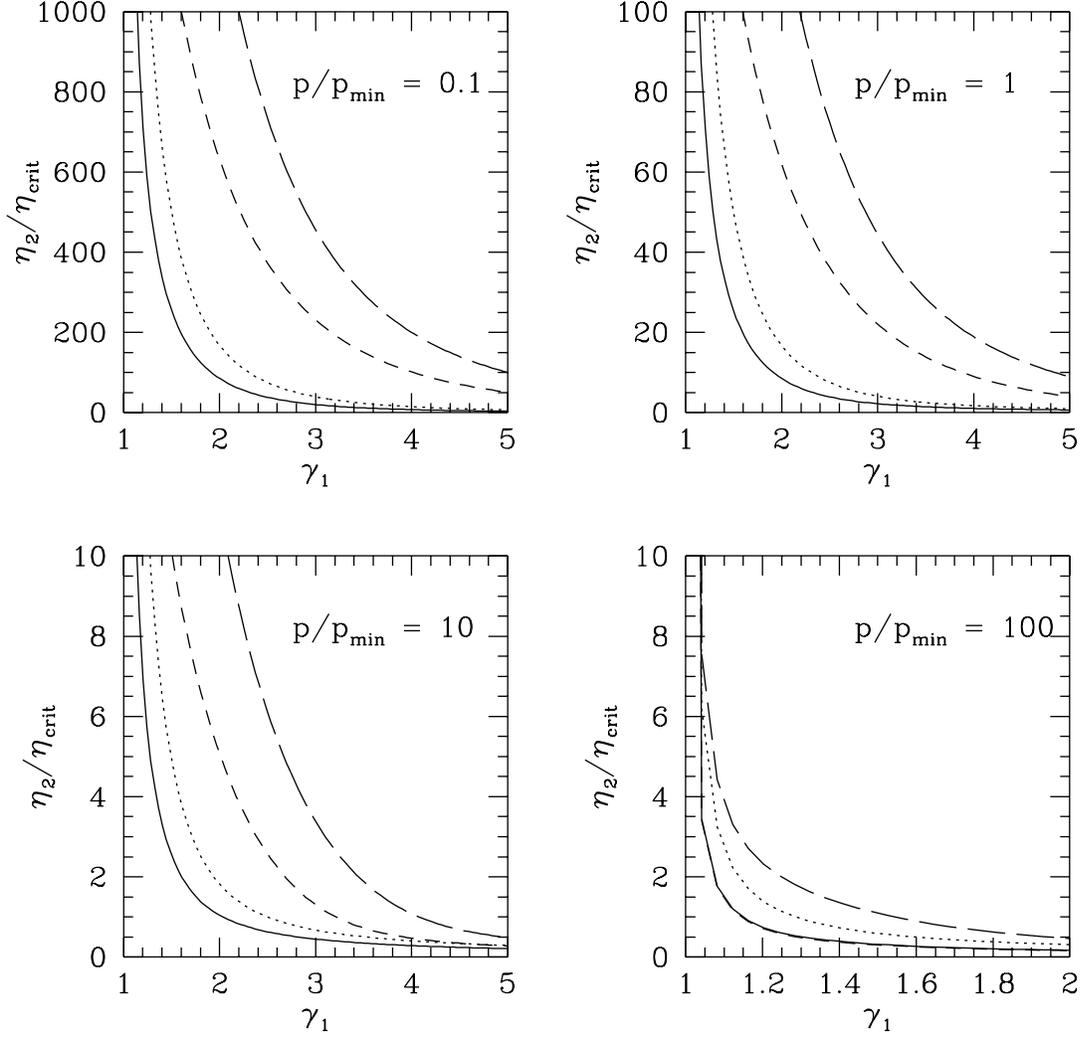

Figure 8: The value of the jet density ratio $\eta$ at 2 different positions along the NGC 315 kpc-scale jet as a function of the pc-scale Lorentz factor for different values of $p/p_{\min}$ and jet inclination, $\theta$. Solid curve: $\Theta = 17''$, $\theta = 90°$, dotted curve: $\Theta = 27''$, $\theta = 30°$, short-dashed curve: $\Theta = 17''$, $\theta = 90°$, long-dashed curve: $\Theta = 27''$, $\theta = 30°$.

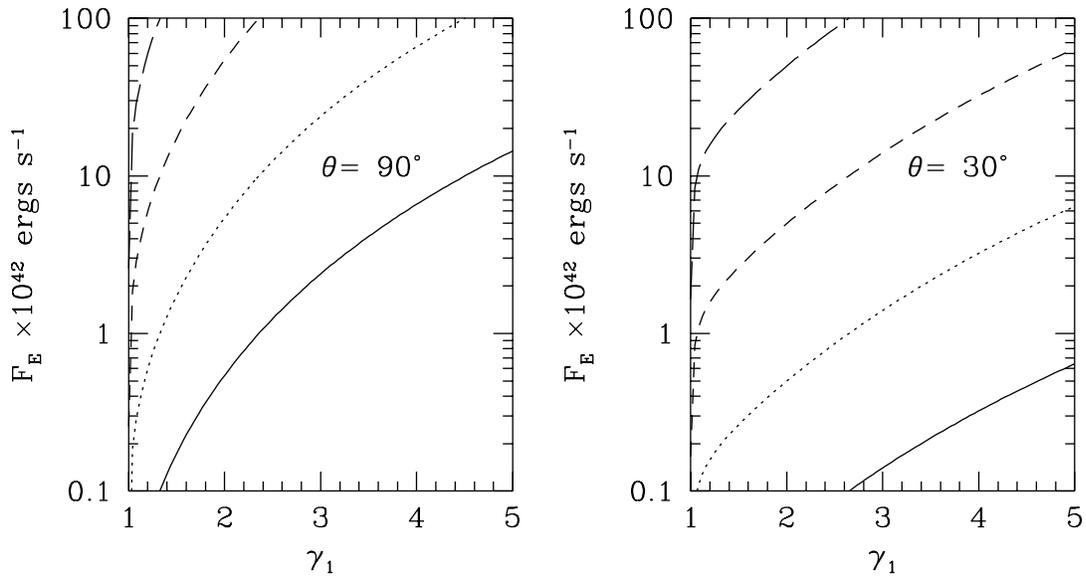

Figure 9: The energy flux in the NGC 315 VLBI jet for different values of $p/p_{\min}$. Solid curve: $p/p_{\min} = 0.1$, dotted curve: $p/p_{\min} = 1$, short-dashed curve: $p/p_{\min} = 10$, long-dashed curve: $p/p_{\min} = 100$. The left hand panel correspond to an inclination of 90°; the right hand panels correspond to an inclination of 30°.

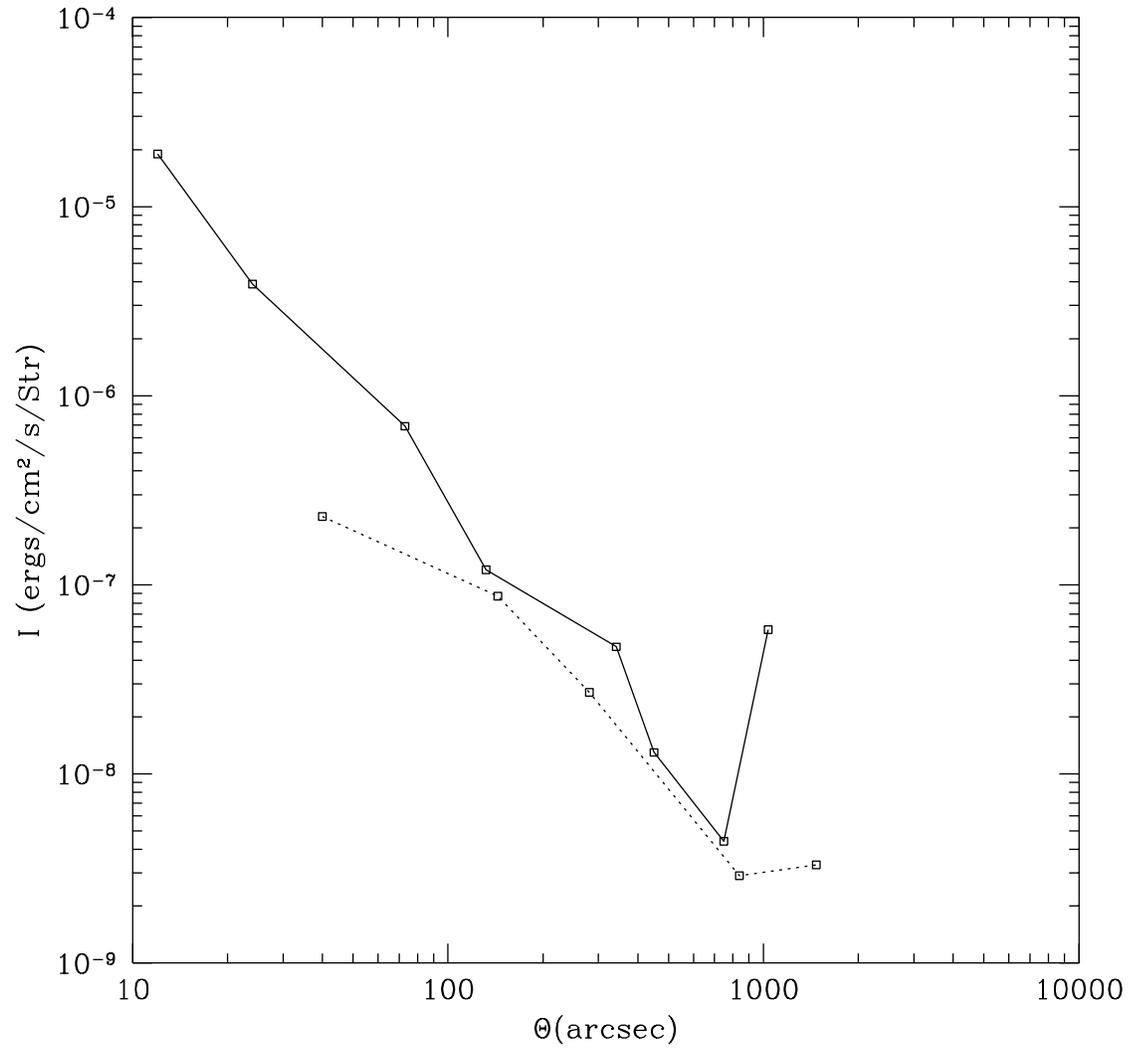

Figure 10: The NGC 315 main jet and counter jet integrated surface brightnesses, from data in Willis *et al.*(1981).

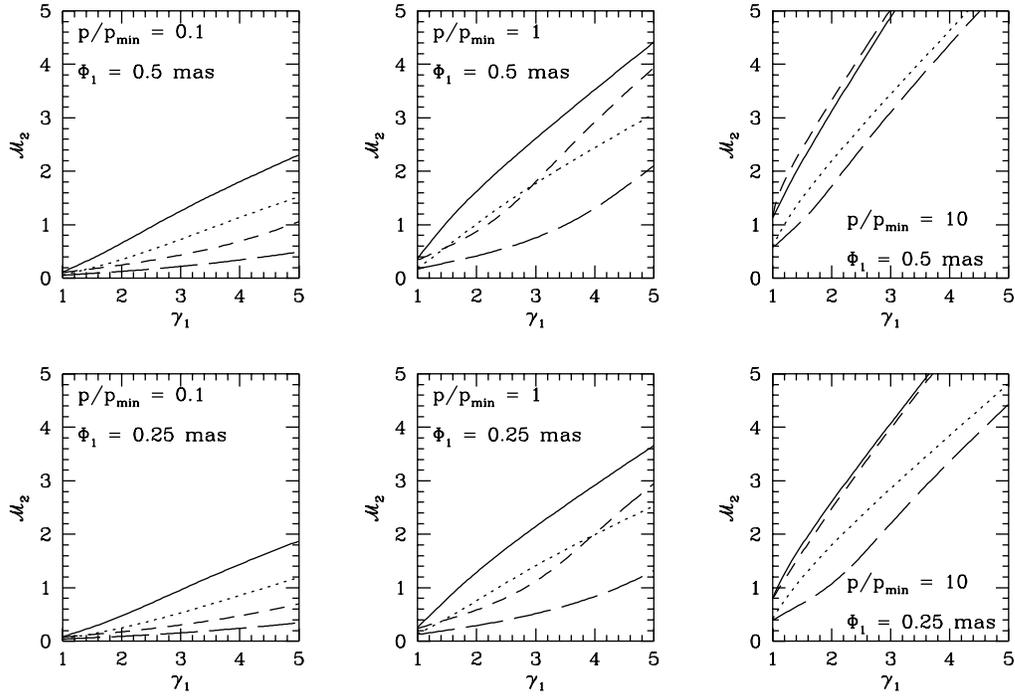

Figure 11: The Mach number at 2 different positions along the NGC 6251 kpc-scale jet as a function of the pc-scale Lorentz factor for different values of $p/p_{\rm min}$, different angles of inclination $\theta$, and different assumed jet diameters. Top panels: $\Phi_1 = 0.5$ mas; lower panels: $\Phi_1 = 0.25$ mas. Solid curve: $\Theta = 10''$, $\theta = 90°$, dotted curve: $\Theta = 20''$, $\theta = 90°$, short-dashed curve: $\Theta = 10''$, $\theta = 30°$, long-dashed curve: $\Theta = 20''$, $\theta = 30°$.

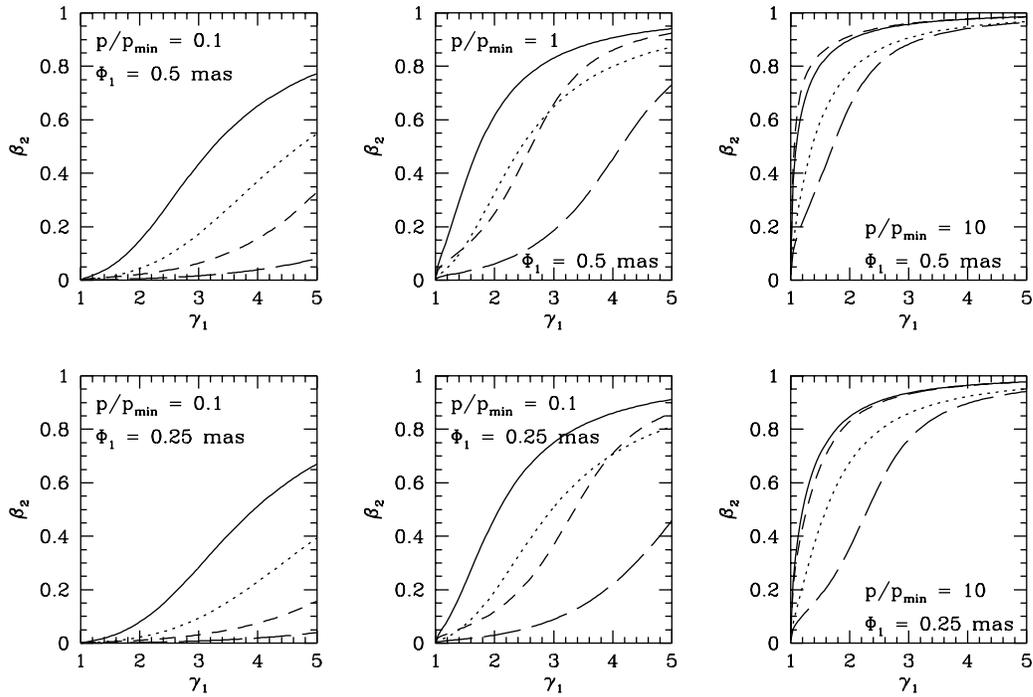

Figure 12: The value of $\beta$ at 2 different positions along the NGC 6251 kpc-scale jet as a function of the pc-scale Lorentz factor for different values of $p/p_{\min}$, different angles of inclination $\theta$, and different assumed jet diameters. Top panels: $\Phi_1 = 0.5$ mas; lower panels: $\Phi_1 = 0.25$ mas. Solid curve: $\Theta = 10''$, $\theta = 90°$, dotted curve: $\Theta = 20''$, $\theta = 90°$, short-dashed curve: $\Theta = 10''$, $\theta = 30°$, long-dashed curve: $\Theta = 20''$, $\theta = 30°$.

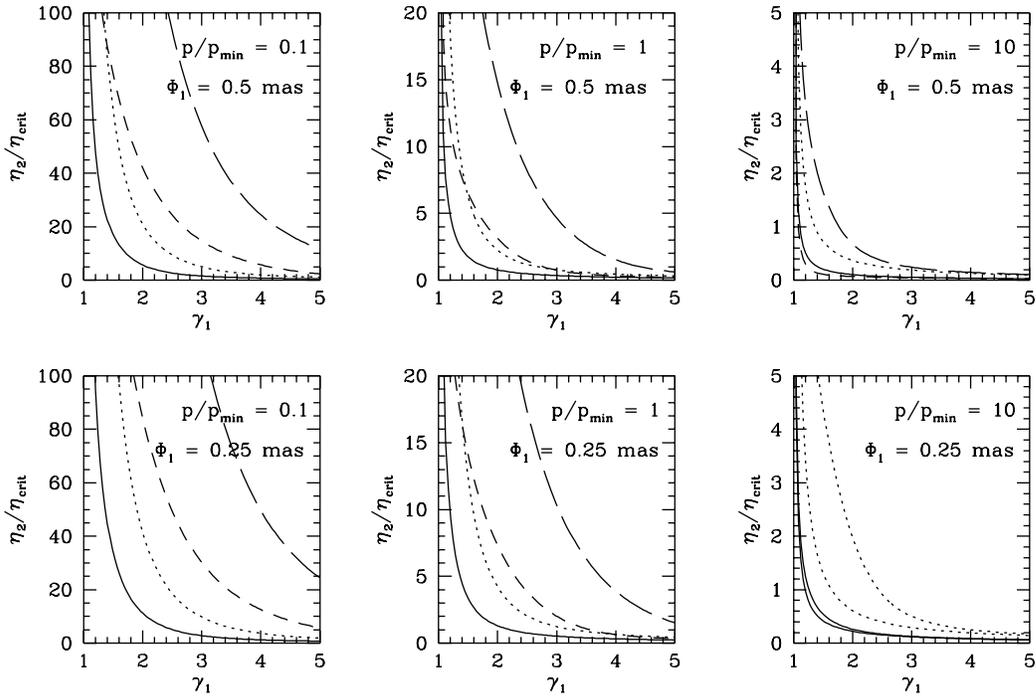

Figure 13: The jet density ratio compared to the critical value ($\eta/\eta_{\rm crit}$) at 2 different positions along the NGC 6251 kpc-scale jet as a function of the pc-scale Lorentz factor for different values of $p/p_{\rm min}$, two different angles of inclination and different assumed jet diameters. Top panels: $\Phi_1 = 0.5$ mas; lower panels: $\Phi_1 = 0.25$ mas. Solid curve: $\Theta = 10''$, $\theta = 90°$, dotted curve: $\Theta = 20'', \theta = 90°$, short-dashed curve: $\Theta = 10''$, $\theta = 30°$, long-dashed curve: $\Theta = 20''$, $\theta = 30°$.

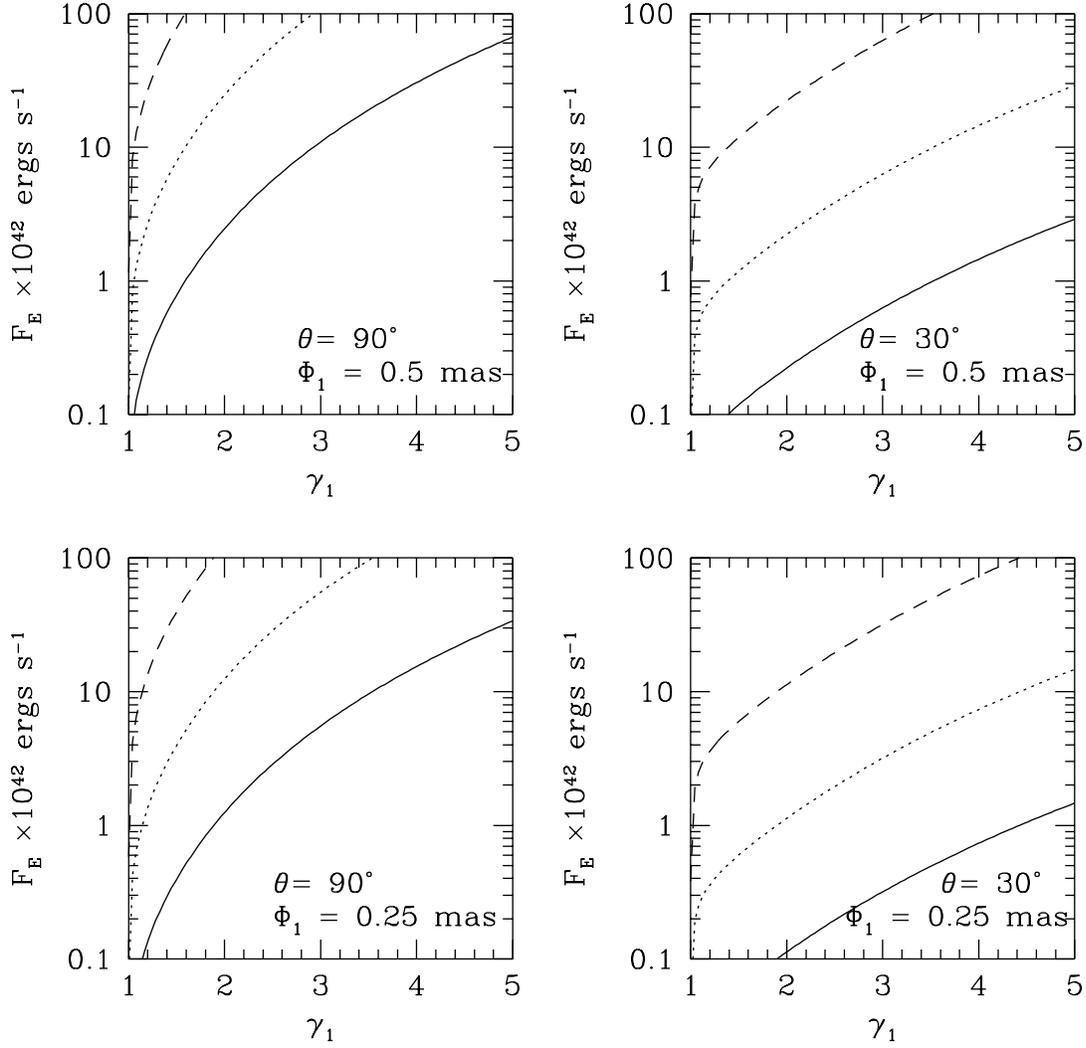

Figure 14: The energy flux in the NGC 6251 VLBI jet as a function of the Lorentz factor. The top panels correspond to an assumed width of 0,5 mas; the lower panels to an assumed width of 0.25 mas. The left-hand panels correspond to a an angle of inclination of 90°; the right hand panels correspond to an angle of inclination of 30°.